\documentclass[11pt]{article}
\usepackage[a4paper,margin=1in]{geometry}
\usepackage{amsmath,amssymb,amsfonts}
\usepackage{booktabs,multirow,array}
\usepackage{graphicx}
\usepackage{microtype}
\usepackage[colorlinks=true,linkcolor=blue,citecolor=blue,urlcolor=blue]{hyperref}
\usepackage{enumitem}
\usepackage{caption}
\usepackage{subcaption}
\usepackage{float}
\graphicspath{{./}{figures/}}
\setlist{nosep}
\title{Contextual Chain: A Controlled Simulation Study of Context-Informed Gossip Scheduling for Post-Partition Recovery}
\author{Song-Ju Kim\\SOBIN Institute LLC, Kawanishi, Hyogo, Japan\\\texttt{kim@sobin.org}}
\date{}

\begin{document}
\maketitle

\begin{abstract}
We introduce \emph{Contextual Chain} as a design principle in which evolving operational context can guide subsequent system decisions, and evaluate a first systems realization for post-partition recovery. The central timing experiment uses explicitly separated common-random-number (CRN) streams so that matched timing policies experience identical proposal, broadcast-channel, and gossip-opportunity traces at exactly the same assigned synchronization budget. Relative to approximately uniform \texttt{FixedMatched} timing, the oracle-assisted \texttt{Adaptive} schedule raises final agreement from 0.794 to 0.832 in Case A and from 0.787 to 0.834 in Case B, shortens mean recovery by 14.4 and 12.1 s, and increases the post-rejoin agreement fraction by about 0.044 in both cases. An offline non-contextual \texttt{DisruptionWindowMatched} control preserves each Adaptive run's exact total budget and exact low/high-rate tick counts while placing high-rate effort deterministically around the known disruption boundary without using the quarantine fraction. This control recovers 2.3 and 1.2 s faster than Adaptive and has essentially the same post-rejoin agreement fraction, while Adaptive retains a smaller final-agreement advantage of 2.3 and 1.8 percentage points. Thus most of the recovery-time benefit is explained by concentrating synchronization around disruption/rejoin rather than by the oracle context signal itself; within this simulator, the contextual trigger contributes a modest additional end-of-horizon agreement benefit. A broader factorial screening study shows that synchronization provisioning is the dominant recovery factor, whereas the Full head-selection rule has no consistent advantage over height-only selection. The evidence is limited to an abstract oracle-assisted simulator and assigned pair opportunities; decentralized estimation, bounded state, packet-level cost, adversarial robustness, and deployment validation remain open.
\end{abstract}


\section{Introduction}
Intermittently connected systems must continue operating while peers disappear, links fail, and different components accumulate incompatible state. This problem appears in disruption-tolerant networking, replicated data, mobile and sensor systems, and ledger-like protocols. Classical epidemic anti-entropy work showed that randomized pairwise exchange can drive replicas toward consistency under weak communication assumptions \cite{Demers1987Epidemic}; disruption-tolerant networking explicitly treats long delays and partitions as architectural conditions rather than exceptions \cite{Cerf2007DTN}. More recent constrained-network work, including SCHC over intermittently available satellite IoT links, further illustrates that recovery cost depends strongly on the communication substrate \cite{MunozLara2026SCHC}.

This paper introduces \textbf{Contextual Chain} at two deliberately separated levels. At the conceptual level, Contextual Chain is a design principle: participants use an evolving, jointly accumulated communication context as operational state. A participant that has remained synchronized needs to follow only its current context, whereas an observer that does not know which history is current may have to preserve several plausible alternatives. We refer to the latter possibility as \emph{context-induced state asymmetry}. It is a research hypothesis about resource asymmetry, not a security theorem.
This framing is informed by prior work on contextual bookkeeping under fixed shared-state reuse and context-sensitive control \cite{KimBookkeeping2026,KimNogo2026,KimControl2026}. Those results provide representational motivation rather than a security proof for the networked resource-asymmetry hypothesis studied here.

Contextual Chain is conceptually distinct from established post-quantum and resource-bounded cryptographic constructions. Modern post-quantum cryptography addresses threats from future quantum computers using standardized mathematical constructions such as ML-KEM, ML-DSA, and SLH-DSA \cite{NISTFIPS203,NISTFIPS204,NISTFIPS205}. Separate cryptographic traditions study security under resource bounds, including bounded quantum storage \cite{Damgard2008BQSM}, bounded-storage authentication \cite{Dodis2022Authentication}, and proofs of space \cite{Dziembowski2015ProofsSpace}. The present study does not instantiate the NIST post-quantum constructions or prove security in a bounded-storage model; these lines of work motivate the broader question of whether shared communication context can create useful resource asymmetry.

At the systems level, we study a first concrete realization: recovery of a ledger-like shared head after a temporary partition. Each simulated node uses a local context-aware branch-acceptance rule, while a synchronization controller increases gossip effort when inconsistency is widespread. The central empirical question is therefore narrower than the long-term authentication motivation: \emph{which part of this first realization actually drives post-rejoin recovery under the simulated conditions?}

The controlled experiments identify the data-supported mechanism more precisely. A CRN equal-budget timing study compares the oracle-assisted \texttt{Adaptive} schedule with two offline matched controls under identical external traces: approximately uniform \texttt{FixedMatched} timing and a non-contextual \texttt{DisruptionWindowMatched} schedule that preserves Adaptive's exact low/high-rate counts while using only the known partition/rejoin boundaries. Adaptive remains clearly better than approximately uniform timing, but the disruption-window control explains most of the recovery-time advantage; the contextual trigger retains only a smaller end-of-horizon agreement benefit. Across a broader mechanism-screening factorial, synchronization \emph{quantity} remains the dominant determinant of recovery, while the Full head-selection rule has no consistent independent advantage over height-only selection.

\paragraph{Contributions.}
\begin{enumerate}
\item We introduce Contextual Chain as a context-based design principle and formulate the \emph{context-induced state asymmetry} hypothesis without claiming cryptographic authentication or post-quantum security.
\item We define a first recovery realization in which evolving operational context informs local branch acceptance and, through an oracle-assisted inconsistency/quarantine signal, adaptive synchronization around a temporary partition.
\item We perform a 6,000-run CRN equal-budget timing study with matched proposal, broadcast-channel, and gossip-opportunity traces, including two offline controls that separate approximately uniform allocation from deterministic disruption-window concentration. The comparison shows that disruption-focused timing explains most of the recovery-speed benefit, while the oracle-derived trigger retains a smaller final-agreement advantage.
\item We use a broader $3\times3$ factorial as a mechanism-screening study of synchronization provisioning and head-selection logic, distinguish assigned synchronization opportunities from packet-level communication cost, and keep the resource-asymmetry extension as exploratory Supplementary motivation rather than cryptographic evidence.
\end{enumerate}

\section{Related work and positioning}
\subsection{Recovery, anti-entropy, and disruption tolerance}
Epidemic replication and anti-entropy protocols explicitly trade communication effort for faster convergence \cite{Demers1987Epidemic}. Delay- and disruption-tolerant networking addresses store-and-forward communication across intermittent contacts \cite{Cerf2007DTN}. Contextual Chain is closest to these works in its recovery objective. The present contribution is a controlled study of how a local branch rule and synchronization provisioning interact after a partition rather than a consistency theorem.
Wireless coordination research also treats synchronization or desynchronization as a controllable systems resource, including DESYNC and synchronization-based resource-allocation methods for constrained wireless networks \cite{Degesys2007DESYNC,Yu2019KuramotoDesynch,YasudaDesynch,KimTOWSync2021}. These methods address medium-access or slot/resource coordination rather than missing-state repair, but they motivate the broader systems view that synchronization effort can be adapted to operating conditions.

\subsection{Ledgers and lightweight verification}
Bitcoin-style longest-chain protocols and Byzantine agreement protocols solve different agreement problems and operate under different assumptions \cite{Nakamoto2008,CastroLiskov1999,Miller2016HoneyBadger}. Other public-ledger consensus families include Algorand and Avalanche \cite{Gilad2017Algorand,Rocket2019Avalanche}; their fault and timing models likewise differ from the post-partition repair setting studied here. NiPoPoWs, FlyClient, and Utreexo reduce verification or retained-state burdens for light participants \cite{Kiayias2020NIPoPoWs,Bunz2019FlyClient,Dryja2019Utreexo}. DAG-oriented ledgers such as the Tangle and IoT-oriented DAG designs pursue different structural, throughput, and security trade-offs \cite{Popov2017Tangle,Li2018DAGIoT}. None of these systems is a direct counterpart of the state-repair scheduling control evaluated here, so an ``identical-condition'' benchmark would require adding a recovery policy around them. We therefore do not claim performance superiority over these systems. Instead, Table~\ref{tab:scope-comparison} separates representative adjacent objectives from the recovery policy studied here.

Blockchain-enabled mobile and IoT communication systems provide additional adjacent context under explicit security, consensus, and application-integration objectives. Chen et al. study practical Byzantine-fault-tolerant robustness for mobile crowdsensing \cite{Chen2023PBFTMCS}; Gul et al. examine blockchain-aided secure communication and game-theoretic coordination in the Internet of Drones \cite{Gul2025IoD}; and Gul surveys blockchain-enabled IoT platforms for vehicular sensing and transportation monitoring \cite{Gul2022BlockchainIoT}. These studies address adversarial robustness, secure coordination, or blockchain integration in mobile/IoT environments. The present work instead isolates non-adversarial post-partition repair scheduling in an abstract simulator and therefore does not claim to reproduce their security models, consensus guarantees, or deployment conditions.

\begin{table}[H]
\centering
\caption{Scope comparison with representative adjacent methods. A dash means that the cited method's primary contribution is not the corresponding recovery function.}
\label{tab:scope-comparison}
\small
\begin{tabular}{>{\raggedright\arraybackslash}p{2.5cm}>{\raggedright\arraybackslash}p{3.1cm}>{\raggedright\arraybackslash}p{2.8cm}>{\raggedright\arraybackslash}p{2.8cm}}
\toprule
Method & Primary objective & Native post-partition recovery policy & Present-paper comparison claim\\
\midrule
NiPoPoW / FlyClient & Succinct chain verification & -- & No direct superiority claim\\
Utreexo & Compact dynamic accumulator / UTXO state & -- & No direct superiority claim\\
Epidemic anti-entropy & Replica convergence by randomized exchange & Pairwise anti-entropy & Conceptually closest synchronization precedent\\
Contextual Chain (this work) & Recovery study with context-aware acceptance plus adaptive gossip & Explicit suffix repair after rejoin & Evaluated only within stated simulator\\
\bottomrule
\end{tabular}
\end{table}

Transaction confirmation-time and reliability models provide another useful comparison point because they emphasize time-to-stability and failure probability rather than only eventual correctness \cite{Malakhov2023Confirmation,Malakhov2024Reliability}. Our recovery-time metric similarly treats delay after rejoin as a first-class outcome, though our system model and mechanism are different.

\subsection{Heterogeneous edge systems and federated learning}
Federated learning also operates across clients with heterogeneous computation and connectivity. Recent work on federated large language models and clustered federated learning highlights the importance of communication-efficient coordination and heterogeneous clients \cite{Cheng2025FedLLM,Cheng2025SnapCFL}. We cite this literature as an adjacent systems motivation, not as a protocol baseline: the objective functions, state semantics, and fault assumptions differ from ledger recovery.

\subsection{Resource-bounded cryptographic motivation}
Bounded-storage and bounded-quantum-storage models demonstrate that cryptographic security can, in some settings, be derived from asymmetric memory capabilities \cite{Damgard2008BQSM,Dodis2022Authentication}. Proofs of space similarly turn storage into a verifiable resource \cite{Dziembowski2015ProofsSpace}. These works are important independent context for our longer-term motivation. The present protocol, however, neither instantiates their formal games nor provides cryptographic soundness. Our Supplementary context-coverage model is therefore described only as a resource-accounting thought experiment.

\section{Contextual Chain design principle}
Let $C_t$ denote the operational context jointly accumulated by synchronized participants up to time $t$. In the simplest idealized interpretation, an honest synchronized participant follows one current trajectory,
\[
C_0 \rightarrow C_1 \rightarrow \cdots \rightarrow C_t.
\]
A context-uncertain observer may instead consider a set $\mathcal{C}_t=\{C_t^{(1)},\ldots,C_t^{(m_t)}\}$ of plausible current contexts. If future acceptance depends on information tied to the true trajectory, the observer may need to preserve several alternatives until subsequent observations eliminate them. The design hypothesis is therefore not that honest devices should consume large memory; it is that shared context may keep the honest state narrow while uncertainty can broaden the alternative-state set elsewhere.

This design hypothesis does not imply a cryptographic guarantee. In the present recovery simulator, ``context'' is realized by a local block graph, current head, recent branch statistics, and quarantine state. The simulator itself retains more state than the ideal conceptual picture; the experiments therefore constitute a first operational exploration and do not demonstrate bounded honest memory.
A versioned public specification documents the broader Contextual Chain design family and Reference Profile R1 separately from the empirical claims tested in this article \cite{KimContextualChainSpec2026}.

\section{Methods}
\subsection{System and network model}
We use the discrete-event simulator described in this study. Unless otherwise stated, the main study uses $N=20$ nodes, a 3,600-s horizon, global mean block interval 30 s, and a hard two-component partition from $t=1200$ s to $t=2400$ s. Case A uses a 50/50 split and Case B an 80/20 split. The main suite uses 1,000 seeds per condition; scaling uses 500 seeds per condition at $N\in\{20,50,100\}$.

The logical network is complete outside the imposed partition. During the partition, cross-component transmission is blocked while ordinary block broadcast can continue within each component; the simulator's periodic repair-gossip routine is suspended until rejoin. Proposal inter-arrival times are exponentially sampled from a fixed global target rate divided evenly among nodes. Delay is sampled from a Gaussian distribution and truncated below at zero. The parameter named \texttt{drop\_p} is applied independently at simulator transmission/delivery stages; it should therefore be read as a per-stage drop parameter rather than a calibrated end-to-end packet-loss probability. The synthetic clean setting uses $(\texttt{drop\_p},\mu_{d},\sigma_d)=(0,0.25,0.10)$ s and the synthetic noisy setting uses $(0.02,0.80,0.20)$ s.

No finite link capacity, queueing model, fragmentation, burst-correlated loss, asymmetric links, node churn, heterogeneous radio rates, or empirical radio traces are included. These conditions are therefore described as a synthetic noisy network rather than as a calibrated representation of a specific IoT deployment. This abstraction supports causal comparison of protocol settings while limiting deployment-level interpretation. The principal simulator settings are summarized in Table~\ref{tab:parameters}.

\begin{table}[H]
\centering
\caption{Principal simulator parameters.}
\label{tab:parameters}
\small
\begin{tabular}{ll}
\toprule
Parameter & Value / interpretation\\
\midrule
Nodes & $N=20$ main; $N=20,50,100$ scaling\\
Simulation horizon & 3,600 s main\\
Global mean block interval & 30 s\\
Partition & 1,200--2,400 s\\
Splits & 50/50 and 80/20 main\\
Decision-score window & $L=40$ blocks\\
Height epoch label & 30 blocks\\
EMA factor & $\alpha=0.85$\\
Quarantine thresholds & $T_{on}=1.05$, $T_{off}=0.75$, off-streak $S=25$\\
Gossip tick & 1 s outside partition\\
Low / high pair budget & 1 / 4 pair selections per tick (main)\\
Synthetic noisy network & per-stage drop parameter 0.02; delay $0.80\pm0.20$ s\\
Seeds & 1,000 main/factorial/equal-budget per cell; 500 scaling\\
\bottomrule
\end{tabular}
\end{table}

\subsection{Context-aware branch acceptance}
Each node stores a block graph with parent pointers, a current head, proposer-indexed reputation values, recent inconsistency statistics, and an orphan pool. In the simulator used here, recent scoring windows are bounded, but the block graph and orphan records are not garbage-collected. Consequently, the complete simulated node state is neither bounded nor compact.

For the main height-based mode, each block is labelled by an epoch level that increments every 30 heights. Because this level is a deterministic monotone function of height when sticky checkpoint-hash tie-breaking is disabled, it is \emph{not an independent fork-choice signal}. The operational ordering is thus more accurately described as an \emph{epoch-labelled height rule} with a recent-history score used only after height ties. For a visible tip $t$,
\[
\mathrm{score}(t)=\sum_{b\in \mathrm{tail}_L(t)}\log\left(1+\max\{0,R[p(b)]\}\right),
\]
where $p(b)$ is the proposer of block $b$ and $L=40$.

The local inconsistency snapshot is
\[
\mathcal I_t=1.2\log(1+\mathrm{forktop})+0.7\sqrt{\mathrm{reorg\_recent}}+0.9\log(1+\mathrm{equiv\_recent}),
\]
with EMA $\bar{\mathcal I}_t=\alpha\bar{\mathcal I}_{t-1}+(1-\alpha)\mathcal I_t$. Quarantine uses hysteresis. Outside quarantine, the node adopts the current rule winner; inside quarantine, head switching is more conservative.

The code contains equivocation detection and reputation penalties/rewards, but the main per-seed data report zero equivocation events in every main run. Thus the present main experiments do not validate Byzantine robustness or the utility of equivocation penalties. Static proposer identifiers are assumed; Sybil resistance, public-key authentication, dynamic identity establishment, selective message-dropping attackers, and malicious block construction are outside scope.

\subsection{Synchronization policy and oracle-assisted trigger}
Periodic gossip samples sender/receiver pairs and sends a missing suffix after finding the highest common ancestor on the sender's chain. This is an idealized application-level repair operation; it is not a packetized transport implementation.

The four principal variants are:
\begin{itemize}
\item \texttt{NoQ}: quarantine disabled; one gossip pair per tick.
\item \texttt{Q\_only}: quarantine enabled; one gossip pair per tick.
\item \texttt{Gossip\_only}: quarantine disabled; four gossip pairs per tick.
\item \texttt{Both}: quarantine enabled; one pair normally and four when the fraction $q$ of quarantined nodes reaches 0.25.
\end{itemize}

A critical implementation assumption is that $q$ is computed exactly from simulator-wide state. The current controller is therefore an \textbf{oracle-assisted experimental controller}, not a deployable decentralized estimator. Its purpose is to test, at the level of assigned synchronization opportunities, whether changing effort at a useful time improves recovery relative to fixed schedules. A deployable variant would need a local or distributed estimate (for example, piggybacked neighborhood quarantine observations) together with delay, staleness, loss, and signalling costs. We do not claim that such a mechanism has been evaluated here.

\subsection{Controlled factorial and common-random-number equal-budget experiments}
To separate head-selection logic from synchronization provisioning, we use a $3\times3$ factorial as a mechanism-screening experiment. The head-selection factor has three levels: \texttt{Full} (epoch-labelled height with recent branch-score tie-breaking), \texttt{HeightOnly} (height with deterministic tie-breaking and no branch-score guidance), and \texttt{BranchScoreOnly} (height plus the recent branch-score tie-break without the epoch label). Under the main height-epoch/sticky-free configuration, the epoch label is a deterministic monotone function of height; therefore \texttt{Full} and \texttt{BranchScoreOnly} are expected to be operationally equivalent. The synchronization factor has three levels: \texttt{FixedLow} (one assigned pair draw per connected gossip tick), \texttt{FixedHigh} (four draws), and \texttt{Adaptive} (one or four according to whether the simulator-wide quarantine fraction $q$ is below or above 0.25). Conservative head switching is disabled in all nine factorial cells. For head-rule comparisons under Adaptive, the same recorded low/high schedule is replayed across head rules.

The common-seed factorial implementation uses one global pseudo-random stream, so shared seed labels do not guarantee lock-step external traces when policies consume different numbers of variates. We therefore use the factorial only for mechanism screening and base the central timing attribution on a separate CRN implementation in which policy-independent stochastic components are separated. Proposal-arrival and block-nonce streams are independent per node; broadcast-channel loss/delay uses a dedicated stream with fixed consumption per proposal-recipient pair; gossip source/destination/drop uses a dedicated stream with fixed consumption per assigned pair opportunity; and variable-length gossip transfer draws are isolated inside each opportunity so that suffix length cannot perturb later opportunities. For every matched case and seed, trace hashes verify equality of proposal, broadcast-channel, and gossip-opportunity external traces across the timing policies used below.

The central timing experiment uses the \texttt{Full} head rule and three schedules. \texttt{Adaptive} is the native oracle-assisted schedule. \texttt{FixedMatched} is an offline control that receives exactly the same integer total assigned pair budget as Adaptive for each case and seed but spreads that budget as uniformly as possible over connected gossip ticks. \texttt{DisruptionWindowMatched} is a stricter offline non-contextual control: it preserves both the exact Adaptive total assigned budget and the exact number of low-rate and high-rate ticks, but places high-rate ticks deterministically immediately after rejoin and, only when necessary to preserve the exact high-rate count, immediately before partition onset. It uses the known partition/rejoin boundaries only and never observes $q$ or any other state-derived context signal. Thus Adaptive versus FixedMatched tests temporal concentration relative to approximately uniform allocation, while Adaptive versus DisruptionWindowMatched tests whether the oracle context signal adds value beyond deterministic knowledge of the disruption boundary.

The CRN timing study uses 1,000 seeds per case and three timing policies, for 6,000 runs in total. Structural validation requires exact budget equality within every case-seed triple and identical proposal, broadcast, and gossip-opportunity trace hashes across all three timing policies before statistical analysis.

\subsection{Metrics and uncertainty analysis}
\textbf{Final agreement} is the fraction of runs whose nodes have the same head at the simulation horizon. In the CRN timing experiment, agreement is additionally sampled on policy-independent 1-s monitor ticks from rejoin to the horizon. \textbf{First sustained recovery} is declared when all heads are equal for 30 consecutive monitor observations; the reported recovery time is the elapsed time from rejoin to the 30th observation completing that sustained window.
This criterion requires 30 consecutive 1-s-spaced agreement observations and is independent of simulator event density.

Durability is reported separately because first sustained recovery does not imply permanent convergence while block proposals continue. \textbf{Post-rejoin agreement fraction} is the fraction of 1-s monitor observations at which all heads are equal. \textbf{Post-recovery divergence} records whether agreement is subsequently lost after the first sustained-recovery event. \textbf{Agreement-loss episodes} counts transitions from all-head agreement to disagreement. \textbf{Final stable agreement time} is the first monitor time after which agreement persists to the simulation horizon; it is defined only for runs that finish in agreement.

The baseline, factorial, and scaling screening analyses use an event-based convergence detector requiring 30 consecutive event-processing checks with a common head. Their recovery values are used only as provisioning/mechanism-screening diagnostics and are not numerically compared with the CRN recovery times. The central timing and durability analyses use only the policy-independent 1-s monitor definition.

In the scaling stress test, some runs never reach the event-based recovery detector by the 3,600-s horizon. These results use administrative right-censoring at 1,200 s after rejoin and RMST$_{1200}$; they serve only as a provisioning stress test and are not part of the CRN timing attribution.

We report 95\% Wilson intervals for binomial final-agreement probabilities where applicable.
The central CRN timing study uses paired 20,000-resample bootstrap intervals for differences in final agreement, recovery time, post-rejoin agreement fraction, and agreement-loss episodes. Paired binary final-agreement comparisons are additionally checked with exact McNemar/binomial tests on discordant seed pairs.
Because the external CRN traces are explicitly verified within each case-seed triple, these timing comparisons support a direct paired common-randomness interpretation; the common-seed factorial is treated only as a screening analysis. We focus interpretation on effect sizes and intervals rather than on isolated $p$ values.

\subsection{Communication accounting and computational complexity}
The simulator records a \texttt{gossip\_pairs} counter, blocks scheduled for gossip/broadcast, and an estimated byte count based on fixed-size block units. The \texttt{gossip\_pairs} counter increments only after a selected pair passes the simulator's initial drop check and therefore measures post-drop-screen pair selections rather than all policy-assigned opportunities. The controlled experiments additionally record a deterministic pre-drop assigned-pair counter that counts the pair opportunities assigned by the synchronization policy. This assigned-pair budget is the quantity matched exactly across Adaptive, FixedMatched, and DisruptionWindowMatched in the CRN timing study. The simulator still does not model transport headers, request/response framing, fragmentation, controller signalling, device radio energy, or realistic retransmission behavior. Equal assigned pair budget therefore does not imply equal packet-level communication cost.

Let $M$ be the number of blocks known to a node, $F$ the number of visible tips, $H$ the longest local chain length, $L$ the recent score window, $G$ the number of sampled gossip pairs per tick, and $K$ the missing suffix length. In the straightforward implementation used here, finding tips scans known blocks, $O(M)$; reconstructing a chain is $O(H)$; fork-choice may evaluate multiple tips and therefore has an implementation-level upper bound $O(M+FH)$, with worst-case quadratic behavior if both $F$ and $H$ scale with $M$. Gossip common-ancestor search is $O(H)$ plus $O(K)$ receiver processing for a suffix. The oracle trigger scans node quarantine state, $O(N)$ per gossip tick. Per-node storage is $O(M+O+P)$ for the unpruned block graph, orphan pool size $O$, and proposer state $P$. These are implementation complexities, not lower bounds; pruning, tip indexing, and cached ancestry would change them substantially.

\section{Results}
\subsection{Main \texorpdfstring{$N=20$}{N=20} noisy study: synchronization quantity strongly affects recovery}
Table~\ref{tab:main-noisy} reports the four baseline variants. The major separation is between the low- and higher-synchronization regimes. In Case A, final agreement rises from 0.581--0.591 for \texttt{NoQ}/\texttt{Q\_only} to 0.837--0.848 for \texttt{Gossip\_only}/\texttt{Both}; Case B shows the same pattern. Recovery means and tails also shorten substantially with more synchronization. These baseline variants establish the provisioning effect but do not by themselves isolate head-selection logic, conservative switching, adaptive timing, and total assigned budget; the controlled experiments below do.

\begin{table}[H]
\centering
\caption{Main synthetic-noisy $N=20$ baseline results (1,000 seeds/condition). Final-agreement values include 95\% Wilson intervals. Recovery values use the original event-based convergence detector and are retained as provisioning diagnostics; mean and $p95$ intervals are 95\% percentile-bootstrap intervals.}
\label{tab:main-noisy}
\scriptsize
\begin{tabular}{llccc}
\toprule
Case & Variant & Final agreement (95\% CI) & Recovery mean s (95\% CI) & Recovery $p95$ s (95\% CI)\\
\midrule
A 50/50 & \texttt{NoQ} & .591 [.560,.621] & 189.9 [185.2,194.8] & 344.1 [326.0,370.1]\\
& \texttt{Q\_only} & .581 [.550,.611] & 194.2 [188.9,199.4] & 365.1 [343.0,375.3]\\
& \texttt{Gossip\_only} & .837 [.813,.859] & 77.8 [75.4,80.2] & 152.1 [142.1,170.0]\\
& \texttt{Both} & .848 [.824,.869] & 82.6 [79.9,85.4] & 172.0 [156.2,182.0]\\
\midrule
B 80/20 & \texttt{NoQ} & .599 [.568,.629] & 167.9 [163.0,172.8] & 329.1 [308.1,346.1]\\
& \texttt{Q\_only} & .607 [.576,.637] & 169.6 [164.7,174.7] & 338.1 [315.1,351.1]\\
& \texttt{Gossip\_only} & .858 [.835,.878] & 69.8 [67.3,72.4] & 154.1 [140.1,163.1]\\
& \texttt{Both} & .847 [.823,.868] & 75.0 [72.2,78.0] & 166.0 [153.0,178.0]\\
\bottomrule
\end{tabular}
\end{table}

Quarantine by itself does not improve the one-pair regime. Conversely, \texttt{Gossip\_only} is comparable to, and on several recovery metrics numerically better than, \texttt{Both}. The confidence intervals reinforce the conservative interpretation: the study establishes a large difference between low and higher synchronization provisioning, but not a recovery superiority of \texttt{Both} over permanently high gossip.

\subsection{Common-seed \texorpdfstring{$3\times3$}{3x3} factorial screens provisioning and head-selection effects}
\begin{figure}[H]
\centering
\includegraphics[width=.96\linewidth]{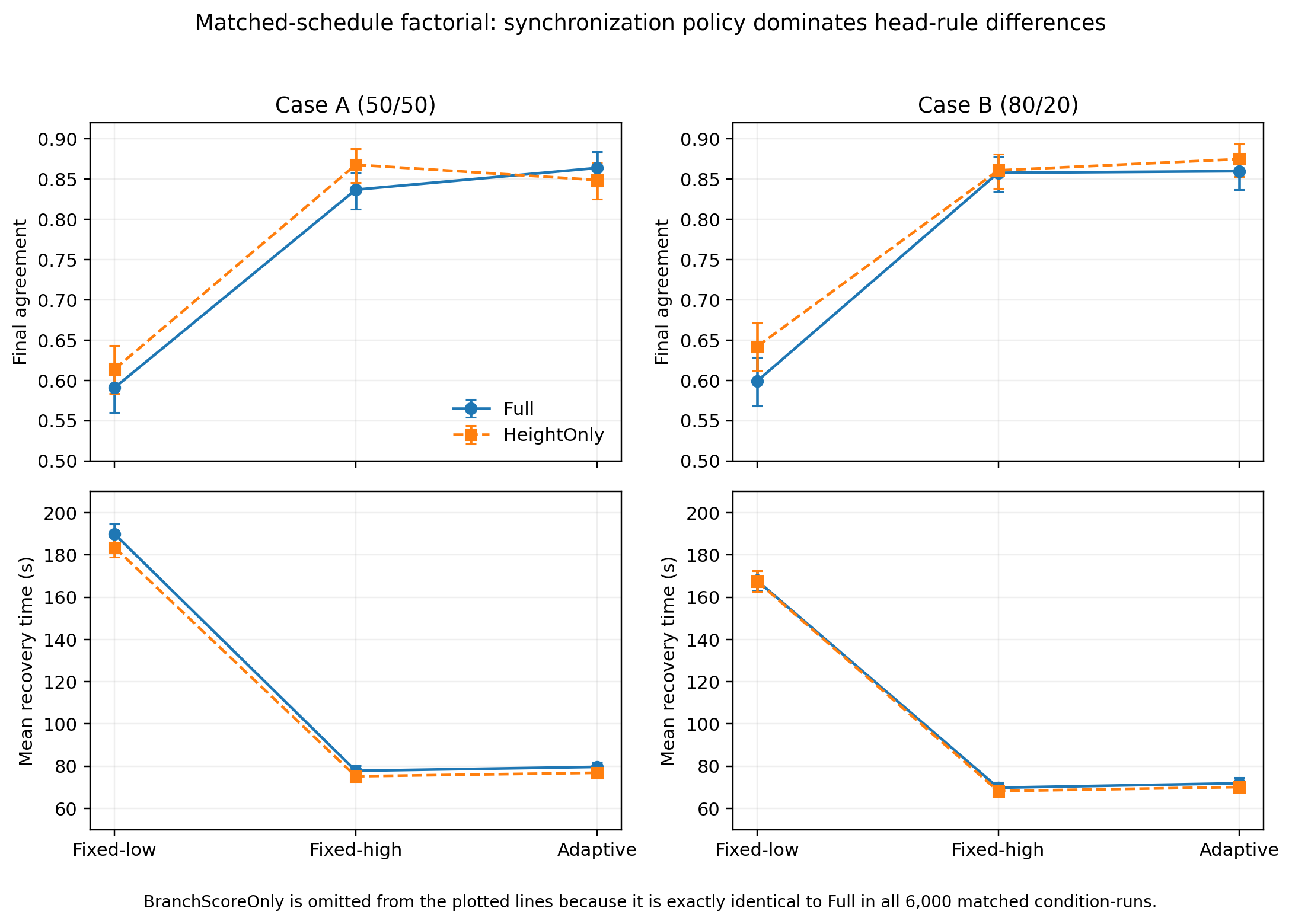}
\caption{Common-seed $3\times3$ factorial screening experiment. Full and HeightOnly are shown; BranchScoreOnly lies exactly on Full in all 6,000 condition-runs. The same Adaptive low/high schedule is replayed across head rules, and conservative head switching is disabled in all factorial cells. Recovery bars use the event-based detector. The dominant descriptive separation is synchronization policy rather than head-selection rule; central timing attribution is based on the CRN experiment in Section~5.3.}
\label{fig:factorial}
\end{figure}

Figure~\ref{fig:factorial} reports the common-seed factorial screening experiment. The synchronization separation is large under both head rules. For the Full rule in Case A, final agreement is 0.591 under FixedLow, 0.837 under FixedHigh, and 0.864 under Adaptive; the event-based recovery mean falls from 189.9 s to 77.8 and 79.7 s. In Case B, the corresponding final-agreement values are 0.599, 0.858, and 0.860, with event-based recovery means 167.9, 69.8, and 71.9 s. Because policy-dependent pseudo-random consumption prevents a lock-step CRN interpretation across synchronization policies, these values are used as mechanism-screening evidence rather than as the central paired timing attribution.

Head-selection effects are much smaller and are not directionally consistent across cases and synchronization policies. HeightOnly is sometimes numerically better and sometimes slightly worse than Full. Most importantly, \texttt{BranchScoreOnly} reproduces \texttt{Full} \emph{exactly} across all 6,000 matched condition-runs for the core metrics. This follows from the deterministic height--epoch relation: under the main height-epoch/sticky-free configuration, the epoch label contributes no independent ordering information. The current data therefore do not support a claim that the fuller head-selection rule itself causes the recovery gain.

\subsection{Common-random-number equal-budget controls separate disruption timing from contextual triggering}
The CRN timing study confirms that the Adaptive advantage over approximately uniform allocation survives explicit CRN control. All 6,000 runs pass the structural checks: every case-seed triple has identical assigned total budget across the three timing policies and identical proposal, broadcast-channel, and gossip-opportunity external trace hashes. Table~\ref{tab:equal-budget-crn} summarizes the cell-level outcomes for the three timing policies in both partition cases.

\begin{table}[H]
\centering
\caption{CRN equal-budget timing study with the Full head rule (1,000 matched seeds per case). All three policies receive the same exact assigned total pair budget for each case and seed. \texttt{DisruptionWindowMatched} also preserves Adaptive's exact low/high-rate tick counts but uses only known partition/rejoin boundaries, not the quarantine fraction.}
\label{tab:equal-budget-crn}
\scriptsize
\resizebox{\textwidth}{!}{%
\begin{tabular}{llcccc}
\toprule
Case & Timing policy & Final agreement & Recovery mean (s) & Post-rejoin agreement fraction & Mean assigned budget\\
\midrule
A 50/50 & Adaptive & .832 & 90.52 & .8278 & 6,810.5\\
& DisruptionWindowMatched & .809 & 88.18 & .8292 & 6,810.5\\
& FixedMatched & .794 & 104.94 & .7832 & 6,810.5\\
\midrule
B 80/20 & Adaptive & .834 & 84.28 & .8332 & 6,813.6\\
& DisruptionWindowMatched & .816 & 83.11 & .8339 & 6,813.6\\
& FixedMatched & .787 & 96.39 & .7889 & 6,813.6\\
\bottomrule
\end{tabular}%
}
\end{table}

Relative to FixedMatched, Adaptive raises final agreement by 3.8 percentage points in Case A (95\% paired-bootstrap interval +2.5 to +5.1 points; exact McNemar $p=1.62\times10^{-9}$) and by 4.7 points in Case B (+3.4 to +6.1; $p=1.18\times10^{-12}$). Mean recovery is 14.43 s faster in Case A (95\% interval $-17.24$ to $-11.75$ s) and 12.11 s faster in Case B ($-14.92$ to $-9.38$ s). The post-rejoin agreement fraction is higher by 0.0445 and 0.0443, respectively. This establishes that temporal allocation matters at fixed assigned quantity under explicitly matched external random traces.

The stricter disruption-window control changes the interpretation of that timing effect. DisruptionWindowMatched itself outperforms FixedMatched in recovery time by 16.76 s in Case A and 13.28 s in Case B, while increasing post-rejoin agreement fraction by 0.0460 and 0.0450. Against this non-contextual disruption-aware baseline, Adaptive retains a smaller final-agreement advantage: +2.3 points in Case A (+1.4 to +3.3; $p=1.55\times10^{-6}$) and +1.8 points in Case B (+0.9 to +2.8; $p=1.21\times10^{-4}$). However, Adaptive reaches first sustained recovery 2.33 s \emph{later} in Case A (+1.27 to +3.52 s) and 1.16 s later in Case B (+0.48 to +1.92 s), and its post-rejoin agreement fraction is nearly the same but slightly lower ($-0.0014$ and $-0.0006$). The data therefore support a narrower conclusion: concentrating equal synchronization effort around a known disruption/rejoin boundary accounts for most of the recovery-time gain over approximately uniform allocation, while the oracle-derived context trigger adds a modest end-of-horizon agreement advantage in this simulator rather than a general recovery-speed advantage.

The exact Adaptive high-rate count cannot always be placed only after rejoin: 49.6\% of matched seeds in each case require an additional pre-partition high-rate window to preserve the exact Adaptive low/high count. This is why DisruptionWindowMatched uses deterministic windows immediately adjacent to both sides of the imposed disruption when necessary.

All runs in the CRN timing study reach the first sustained-agreement event defined by 30 consecutive one-second observations, but every run later loses all-head agreement at least once while proposals continue. Mean agreement-loss episodes range from 30.9 to 32.8 across the six case-policy cells. Final stable agreement is therefore a distinct endpoint: it exists only in runs that finish in agreement (rates 0.787--0.834), and among those runs its mean start time is approximately 1,169 s after rejoin, close to the horizon. These diagnostics confirm that ``first sustained recovery'' is an event-time measure, not permanent convergence.

\subsection{Synchronization budget is not packet-level communication cost}
The controlled studies count policy-assigned pair opportunities before the simulator's initial drop screen. In the broader factorial, FixedLow assigns 2,400 pair draws per run and FixedHigh 9,600. In the CRN timing study, the matched mean assigned budgets are 6,810.5 in Case A and 6,813.6 in Case B, with exact equality across Adaptive, FixedMatched, and DisruptionWindowMatched within every case and seed. This is a policy-level allocation variable, not a traffic measurement.

The simulator also records a post-drop-screen \texttt{gossip\_pairs} counter and blocks scheduled for gossip/broadcast, but it does not model transport headers, request/response framing, fragmentation, controller signalling, finite-capacity queues, realistic retransmissions, or device radio energy. Equal assigned pair budget therefore does not imply equal packet count, byte count, airtime, or energy. The present paper makes no practical communication-efficiency or energy-saving claim from the assigned-budget comparison.

\subsection{Scaling with a fixed total budget is a provisioning stress test}
Table~\ref{tab:scaling} presents the fixed-total-budget scaling experiment as a provisioning stress test. Holding total pair selections fixed as $N$ grows lowers synchronization effort per node, so the result is not a clean measure of intrinsic algorithmic scalability. This is especially clear at $N=50$ and $N=100$.

\begin{table}[H]
\centering
\caption{Case A synthetic-noisy scaling provisioning stress test (500 seeds/condition). Recovery uses the original event-based detector. Final agreement and recovery-by-horizon have 95\% Wilson intervals; RMST$_{1200}$ treats unrecovered runs as right-censored at 1,200 s after rejoin and reports a 95\% bootstrap interval.}
\label{tab:scaling}
\scriptsize
\begin{tabular}{clccc}
\toprule
$N$ & Variant & Final agreement (95\% CI) & Recovery by horizon (95\% CI) & RMST$_{1200}$ s (95\% CI)\\
\midrule
20 & \texttt{NoQ} & .594 [.550,.636] & 1.000 [.992,1.000] & 194.3 [187.4,201.7]\\
20 & \texttt{Q\_only} & .620 [.577,.661] & 1.000 [.992,1.000] & 195.0 [187.8,202.3]\\
20 & \texttt{Gossip\_only} & .846 [.812,.875] & 1.000 [.992,1.000] & 75.9 [72.7,79.4]\\
20 & \texttt{Both} & .856 [.823,.884] & 1.000 [.992,1.000] & 82.5 [78.5,86.8]\\
\midrule
50 & \texttt{NoQ} & .094 [.071,.123] & .690 [.648,.729] & 842.9 [816.2,870.0]\\
50 & \texttt{Q\_only} & .078 [.058,.105] & .660 [.617,.700] & 848.6 [821.5,877.1]\\
50 & \texttt{Gossip\_only} & .558 [.514,.601] & 1.000 [.992,1.000] & 178.5 [171.1,185.8]\\
50 & \texttt{Both} & .514 [.470,.558] & 1.000 [.992,1.000] & 195.3 [187.0,204.2]\\
\midrule
100 & \texttt{NoQ} & .002 [.000,.011] & .010 [.004,.023] & 1196.7 [1193.0,1199.5]\\
100 & \texttt{Q\_only} & .002 [.000,.011] & .014 [.007,.029] & 1196.7 [1193.5,1199.3]\\
100 & \texttt{Gossip\_only} & .162 [.132,.197] & .932 [.906,.951] & 548.0 [520.7,575.8]\\
100 & \texttt{Both} & .194 [.162,.231] & .918 [.891,.939] & 554.7 [528.1,582.1]\\
\bottomrule
\end{tabular}
\end{table}

The $N=100$ results remain poor in final agreement even for the higher-synchronization variants, and this paper does not provide a solution at that size. Future scalability claims require normalized budgets, topology-aware peer selection, cluster/relay structures, and finite-capacity links.

\subsection{\texorpdfstring{$N=50$}{N=50} budget sweep}
A separate $N=50$ sweep increases fixed and adaptive pair budgets. Representative operating points are shown in Table~\ref{tab:n50-budget}. Raising the budget moves both fixed and adaptive policies into a much better regime. The result supports a provisioning interpretation of the observed $N=50$ degradation, but it does not establish a general scaling law.

\begin{table}[H]
\centering
\caption{Representative $N=50$ noisy budget-sweep operating points. Recovery $p95$ uses the original event-based detector. The final column is the post-drop-screen gossip-pair counter, not assigned pair budget or packet-level byte cost.}
\label{tab:n50-budget}
\scriptsize
\begin{tabular}{llccc}
\toprule
Case & Variant & Final agreement & Recovery $p95$ (s) & Mean post-drop-screen gossip-pair counter\\
\midrule
A & \texttt{Both\_1\_12} & .750 & 193.05 & 17,672\\
A & \texttt{Both\_1\_16} & .772 & 173.05 & 23,242\\
A & \texttt{Gossip\_only\_16\_16} & .798 & 172.25 & 37,634\\
B & \texttt{Both\_1\_12} & .770 & 190.05 & 17,705\\
B & \texttt{Both\_1\_16} & .810 & 157.20 & 23,287\\
B & \texttt{Gossip\_only\_16\_16} & .808 & 137.05 & 37,633\\
\bottomrule
\end{tabular}
\end{table}

The budget sweep remains a provisioning diagnostic. Because its post-drop-screen pair counter is not the same quantity as the pre-drop assigned-pair budget used in the controlled experiments, and because controller signalling and packet costs are omitted, it is not used for the equal-budget timing claim above.

\section{Discussion}
\subsection{The demonstrated mechanism in the first Contextual Chain realization}
The CRN study supports two distinct conclusions. First, temporal concentration matters at fixed assigned synchronization quantity: Adaptive remains substantially better than approximately uniform FixedMatched timing under identical matched external traces. Second, the contextual quarantine signal is not responsible for most of the recovery-speed gain. The offline deterministic DisruptionWindowMatched control uses only the known disruption boundary, while preserving Adaptive's exact total budget and exact low/high-rate count; it recovers slightly faster than Adaptive and achieves essentially the same post-rejoin agreement fraction. Adaptive retains a statistically supported but much smaller final-agreement advantage of 1.8--2.3 percentage points.

The empirically supported mechanism is therefore best described as \emph{disruption-focused temporal allocation of synchronization effort}, with a modest additional endpoint effect from the oracle-derived context trigger under the tested simulator. The data do not establish that the quarantine fraction is generally superior to other adaptive or event-triggered scheduling signals. Across the broader factorial, synchronization quantity remains the dominant determinant of recovery, and the present fuller head-selection rule is not the source of the timing effect: HeightOnly is comparable to Full, while BranchScoreOnly is operationally equivalent to Full in the main configuration.

This interpretation also clarifies the role of the oracle. The exact simulator-wide quarantine fraction remains useful as an experimental upper-bound signal for asking whether state-dependent timing can matter, but it is not a deployable distributed controller. A practical estimator could be delayed, lossy, stale, biased by local topology, and costly to signal. The present result therefore does not imply that a decentralized implementation will reproduce the same endpoint advantage.

\subsection{Exploratory context-induced state asymmetry}
The longer-term resource-asymmetry motivation remains separate from the demonstrated recovery result. Supplementary Information contains a budget-sensitive context-coverage resource model with an exact combinatorial interpretation. The model shows how uncertainty can expand the number of contextual alternatives that must be covered under its stated challenge rule, but it is not a cryptographic proof, an optimal-adversary result, or evidence of a formal memory lower bound. A cryptographic follow-up would require an explicit security game, concrete commitments, adversarial strategies, and a soundness proof.

\subsection{Deployment mapping to IoT technologies}
The current simulator is above the PHY/MAC layer, so LoRa/LoRaWAN, Zigbee, Bluetooth Low Energy, and Sigfox cannot be treated as interchangeable parameter labels. Qualitatively, short-range Zigbee/BLE deployments tend to offer different payload, duty-cycle, topology, and latency constraints from long-range LPWAN technologies such as LoRaWAN or Sigfox. A deployment study should map each technology to measured contact patterns, MTU/fragmentation behavior, throughput, duty-cycle restrictions, asymmetric loss, and energy per transmission. The SCHC satellite-IoT study is directly relevant because it shows how fragmentation and acknowledgement choices alter delay and energy under intermittent LPWAN connectivity \cite{MunozLara2026SCHC}. Contextual Chain would sit above such a calibrated transport; the present paper does not claim that its synthetic settings predict performance on any one radio technology.

\subsection{Hardware/software co-design}
A future implementation could offload hashing, coding, compression, or recovery primitives to accelerators, but the bottleneck identified here is first a communication-provisioning problem. Hardware acceleration would not by itself repair missing information. Hardware/software co-design is therefore a deployment direction rather than evidence for the present protocol.

\subsection{Limitations}
The present study remains deliberately narrow. The Adaptive trigger uses a simulator-wide oracle rather than a distributed estimator. The network is a complete logical graph outside the imposed partition and uses synthetic delay and independent simulator-stage drops; it omits finite-capacity links, queueing, fragmentation, burst/correlated loss, asymmetric links, churn, heterogeneous rates, and empirical radio/contact traces. Assigned pair opportunities are not packet-level traffic or energy. The local block graph and orphan state are unpruned, so the current implementation does not demonstrate bounded node memory. The main threat model has no malicious proposers, Sybil resistance, cryptographic authentication, or Byzantine security guarantee. The fixed-total-budget $N=100$ result remains poor and is a provisioning stress test rather than evidence of practical scalability. Finally, the broader context-coverage model in Supplementary Information is a non-cryptographic resource-accounting model, not a security proof.

The CRN timing study also has a specific comparison boundary. It establishes an advantage over approximately uniform timing and compares the oracle trigger with one deterministic disruption-boundary baseline, but it does not establish superiority over the broader class of non-uniform, local-state, contact-aware, epidemic-repair, or event-triggered schedulers. Controller sensitivity also remains incomplete: the CRN study uses the default quarantine threshold 0.25, EMA factor 0.85, hysteresis settings 1.05/0.75, off-streak length 25, low/high rates 1/4, and the fixed 1,200-s partition rather than jointly sweeping these control and impairment parameters. Because all CRN runs later leave all-head agreement at least once under continuing proposals, first sustained recovery must not be interpreted as permanent convergence; final agreement and post-rejoin agreement fraction provide complementary durability information.

\subsection{Future work}
The first priority is to replace the simulator-wide quarantine fraction with locally observable or neighborhood-aggregated estimators and test them under delayed, lossy, stale, and topology-biased observations while accounting for their signalling overhead. This should include stronger non-contextual and adaptive baselines, such as randomized post-rejoin bursts, time-since-rejoin policies, missing-suffix-volume triggers, locally observed disagreement, and established adaptive anti-entropy or contact-aware scheduling where technically comparable. A systematic controller study should additionally sweep the quarantine threshold, EMA and hysteresis settings, off-streak length, low/high synchronization rates, partition duration, and loss severity rather than relying on the present default operating point.

A second direction is deployment realism. Trace-driven, packet-level, or emulated evaluation should introduce finite link rates, variable suffix sizes, explicit request/response framing, packetization and fragmentation, retransmissions, correlated/asymmetric loss, churn, and device-level airtime or energy measurements. A third direction is scalable state and communication management: pruning/garbage collection with retained-ancestor invariants, indexed ancestry/tip structures, normalized per-node or per-edge synchronization budgets, and sparse, hierarchical, relay-based, or topology-aware peer selection should be evaluated beyond $N=50$ and specifically against the unresolved $N=100$ regime. Finally, any future security claim should be developed separately through explicit identities, adversary capabilities, a formal security game, concrete commitments/authentication, and proofs; the present context-coverage motivation is insufficient for such a claim.

\section{Conclusion}
The controlled study supports a focused result for the first Contextual Chain recovery realization. With explicit common-random-number matching and exactly equal assigned synchronization budgets, the oracle-assisted Adaptive schedule is clearly better than approximately uniform FixedMatched timing: final agreement improves by 3.8--4.7 percentage points, mean first sustained recovery is 12.1--14.4 s faster, and the post-rejoin agreement fraction increases by about 0.044. However, the offline non-contextual DisruptionWindowMatched control, which preserves the same total budget and the same low/high-rate tick counts while using only known disruption boundaries, explains most of the recovery-time advantage. It recovers 1.2--2.3 s faster than Adaptive with essentially the same agreement fraction, while Adaptive retains a smaller 1.8--2.3-point final-agreement advantage.

The principal systems conclusion is therefore not that the present global context signal is a generally superior adaptive trigger. Rather, synchronization quantity is the dominant recovery resource, and at fixed quantity, concentrating that effort around disruption/rejoin is substantially more effective than approximately uniform allocation. The oracle-derived quarantine signal provides a modest additional end-of-horizon agreement benefit under the tested conditions, but its deployable value remains unestablished. The fuller head-selection rule likewise has no consistent advantage over HeightOnly in the broader factorial. The evidence remains confined to an abstract non-adversarial simulator and assigned-pair accounting; distributed estimation, stronger adaptive baselines, bounded-state implementation, packet-level and trace-driven validation, normalized scalability, and formal security analysis remain future work.

\section*{Data and code availability}
The reproducibility package contains the simulator used for the screening analyses, the separated-stream CRN simulator, CRN trace-validation script, three-policy equal-budget runner, all 6,000 per-run outputs, matched schedules, paired statistical tables, provenance, SHA-256 checksums, and reproduction instructions. The tagged software release (\texttt{v2.1-minor-revision-2026-09-06}) is archived on Zenodo at DOI 10.5281/zenodo.22523507. The archived release contains the CRN materials used for the central analyses reported here and permits regeneration of the reported statistics from the per-seed outputs.

\section*{Author contributions}
S.-J. Kim conceived the Contextual Chain concept, designed the simulator and experiments, analysed and interpreted the results, and prepared the manuscript.

\section*{Competing interests}
The author declares no competing interests.

\section*{Use of generative AI tools}
OpenAI ChatGPT was used as an author-directed tool for language and structural editing and coding assistance. The author independently reviewed and verified all code changes, analyses, interpretations, citations, and final text and takes full responsibility for the manuscript.

\bibliographystyle{unsrt}
\bibliography{Refs}

\clearpage
\begin{center}
{\Large\bfseries Supplementary Information}\\[0.5em]
{\large Contextual Chain: A Controlled Simulation Study of Context-Informed Gossip Scheduling for Post-Partition Recovery}\\[0.5em]
Song-Ju Kim
\end{center}
\vspace{1em}

\setcounter{section}{0}
\setcounter{subsection}{0}
\setcounter{table}{0}
\setcounter{figure}{0}
\renewcommand{\thesection}{S\arabic{section}}
\renewcommand{\thesubsection}{S\arabic{section}.\arabic{subsection}}
\renewcommand{\thetable}{S\arabic{table}}
\renewcommand{\thefigure}{S\arabic{figure}}
\renewcommand{\theHsection}{S\arabic{section}}
\renewcommand{\theHsubsection}{S\arabic{section}.\arabic{subsection}}
\renewcommand{\theHtable}{S\arabic{table}}
\renewcommand{\theHfigure}{S\arabic{figure}}

\section{Purpose and interpretation}
This Supplementary Information provides detailed controls, robustness checks, and exploratory analyses supporting the main Article. The central equal-budget experiment uses explicitly separated common-random-number (CRN) streams and compares the oracle-assisted \texttt{Adaptive} schedule with two offline matched controls: approximately uniform \texttt{FixedMatched} timing and a non-contextual \texttt{DisruptionWindowMatched} schedule. The comparison shows that concentrating synchronization effort around disruption/rejoin accounts for most of the recovery-time gain, while the oracle-derived contextual trigger retains a smaller end-of-horizon agreement advantage. The broader factorial shows that synchronization quantity is the dominant recovery factor and that the Full head-selection rule does not provide a consistent independent recovery advantage. The context-coverage model in Section~\ref{sec:coverage} is an exploratory resource model, not a cryptographic authentication protocol or security proof.

\section{Synchronization counters and accounting limitations}
The simulator records a \texttt{gossip\_pairs} counter and scheduled block transfers. The \texttt{gossip\_pairs} counter increments only after a selected sender/receiver pair passes the simulator's initial drop check and is therefore a post-drop-screen counter rather than the exact number of policy-assigned pair opportunities. The factorial and matched-budget analyses additionally use \texttt{gossip\_budget\_pair\_draws}, which counts assigned opportunities before drop. The fixed-block byte estimate excludes packet headers, requests, fragmentation, controller signalling, and realistic retransmissions; it is not a network-byte measurement. Table~\ref{tab:synchronization-proxies} summarizes these synchronization and block-derived accounting proxies for the main noisy cases.

\begin{table}[H]
\centering
\caption{Main synthetic-noisy $N=20$ synchronization proxies. Estimated bytes use the simulator's fixed per-block accounting.}
\label{tab:synchronization-proxies}
\scriptsize
\begin{tabular}{llrrrr}
\toprule
Case & Variant & Final agreement & Mean post-drop pair counter & Mean gossip blocks & Est. total KiB\\
\midrule
A 50/50 & \texttt{NoQ} & .591 & 2352.0 & 485.7 & 568.4\\
& \texttt{Q\_only} & .581 & 2352.2 & 489.3 & 569.5\\
& \texttt{Gossip\_only} & .837 & 9408.4 & 508.1 & 572.7\\
& \texttt{Both} & .848 & 6711.3 & 503.0 & 574.4\\
\midrule
B 80/20 & \texttt{NoQ} & .599 & 2352.0 & 361.2 & 573.0\\
& \texttt{Q\_only} & .607 & 2352.1 & 360.9 & 572.9\\
& \texttt{Gossip\_only} & .858 & 9408.3 & 379.1 & 576.8\\
& \texttt{Both} & .847 & 6713.9 & 371.6 & 574.2\\
\bottomrule
\end{tabular}
\end{table}

The large change in the post-drop-screen pair counter but small change in block-derived bytes illustrates why neither quantity alone can support a claim of real traffic savings. A deployable evaluation must log assigned attempts, application requests/responses, empty exchanges, control bytes, duplicates, fragmentation, retries, and radio-level energy.

\section{Matched-schedule \texorpdfstring{$3\times3$}{3x3} factorial: full numerical results}
The factorial uses 1,000 common seed labels per cell. Conservative head switching is disabled in all cells. For each case and seed, the Adaptive low/high schedule is generated once by Full+Adaptive and replayed unchanged across Full, HeightOnly, and BranchScoreOnly. FixedLow assigns 2,400 pair draws and FixedHigh 9,600 per run. This common-seed implementation uses one mutable pseudo-random stream, so a shared seed label does not guarantee lock-step proposal, delay, loss, and pair-selection traces when policies consume different numbers of random variates. The factorial is therefore used as a provisioning/head-rule screening analysis rather than as the strongest paired causal timing comparison. Table~\ref{tab:factorial-full} reports the cell-level estimates.

\begin{table}[H]
\centering
\caption{Matched-seed factorial results. Agreement intervals are 95\% Wilson intervals; recovery-mean and $p95$ intervals are 95\% bootstrap intervals. Assigned budget is the pre-drop policy counter.}
\label{tab:factorial-full}
\scriptsize
\resizebox{\textwidth}{!}{%
\begin{tabular}{lllcccc}
\toprule
Case & Head & Synchronization & Final agreement (95\% CI) & Recovery mean s (95\% CI) & Recovery $p95$ s (95\% CI) & Mean assigned pair budget\\
\midrule
A & Full & FixedLow & .591 [.560,.621] & 189.9 [185.0,194.8] & 344.1 [326.0,370.1] & 2400.0\\
A & Full & FixedHigh & .837 [.813,.859] & 77.8 [75.5,80.2] & 152.1 [142.2,168.1] & 9600.0\\
A & Full & Adaptive & .864 [.841,.884] & 79.7 [77.3,82.0] & 153.0 [142.1,170.1] & 6847.8\\
A & HeightOnly & FixedLow & .614 [.583,.644] & 183.5 [178.8,188.5] & 331.0 [315.1,351.0] & 2400.0\\
A & HeightOnly & FixedHigh & .868 [.846,.888] & 75.2 [73.0,77.4] & 146.0 [137.0,157.0] & 9600.0\\
A & HeightOnly & Adaptive & .849 [.825,.870] & 76.9 [74.6,79.3] & 146.1 [138.1,156.0] & 6847.8\\
A & BranchScoreOnly & FixedLow & .591 [.560,.621] & 189.9 [185.2,194.8] & 344.1 [326.0,370.1] & 2400.0\\
A & BranchScoreOnly & FixedHigh & .837 [.813,.859] & 77.8 [75.4,80.2] & 152.1 [142.1,168.1] & 9600.0\\
A & BranchScoreOnly & Adaptive & .864 [.841,.884] & 79.7 [77.3,82.1] & 153.0 [142.1,170.1] & 6847.8\\
\midrule
B & Full & FixedLow & .599 [.568,.629] & 167.9 [163.0,172.6] & 329.1 [310.0,346.1] & 2400.0\\
B & Full & FixedHigh & .858 [.835,.878] & 69.8 [67.3,72.2] & 154.1 [140.1,163.1] & 9600.0\\
B & Full & Adaptive & .860 [.837,.880] & 71.9 [69.5,74.5] & 154.0 [142.1,162.0] & 6850.2\\
B & HeightOnly & FixedLow & .642 [.612,.671] & 167.5 [162.7,172.5] & 336.1 [317.0,344.1] & 2400.0\\
B & HeightOnly & FixedHigh & .861 [.838,.881] & 68.2 [65.9,70.5] & 142.1 [130.0,155.0] & 9600.0\\
B & HeightOnly & Adaptive & .875 [.853,.894] & 70.1 [67.7,72.5] & 145.1 [134.0,156.2] & 6850.2\\
B & BranchScoreOnly & FixedLow & .599 [.568,.629] & 167.9 [162.8,172.9] & 329.1 [308.1,346.1] & 2400.0\\
B & BranchScoreOnly & FixedHigh & .858 [.835,.878] & 69.8 [67.3,72.3] & 154.1 [140.1,163.1] & 9600.0\\
B & BranchScoreOnly & Adaptive & .860 [.837,.880] & 71.9 [69.4,74.4] & 154.0 [142.1,161.1] & 6850.2\\
\bottomrule
\end{tabular}%
}
\end{table}

Full and BranchScoreOnly are exactly identical across all 6,000 common-seed condition-runs for the audited core outcomes. This is not a sampling coincidence: with height-based epoch labels and no sticky checkpoint identity, the epoch label is a deterministic monotone function of height and adds no independent ordering information. HeightOnly differs, but its effects are small and inconsistent compared with the much larger synchronization-provisioning effect. Because the common-seed factorial does not provide lock-step CRN matching across timing policies, it is used to screen the relative scale of provisioning and head-rule effects; the separated-stream CRN experiment below provides the primary timing attribution.

\section{Common-random-number equal-budget timing controls}
\label{sec:crn-equal-budget}
The central timing experiment uses a separated-stream CRN implementation. In the common-seed screening implementation, proposal arrivals, block nonces, broadcast delay/drop outcomes, and gossip-pair selection share one mutable pseudo-random stream. Policies that schedule different numbers of gossip opportunities can therefore consume different numbers of variates and subsequently experience different nominally ``external'' realizations even under the same seed label. The CRN implementation separates proposal, block, broadcast-channel, and gossip-opportunity streams so that the compared timing policies receive identical exogenous traces wherever the experimental design permits.

For each case and seed, the Full+Adaptive run first determines the exact assigned synchronization budget and the exact number of low/high-rate connected ticks. Two offline controls are then constructed. \texttt{FixedMatched} preserves the exact total assigned budget but distributes the extra opportunities approximately uniformly over connected gossip ticks. \texttt{DisruptionWindowMatched} preserves both the exact total budget and the exact low/high-rate tick counts but does not use the quarantine fraction: it places high-rate ticks deterministically adjacent to the known disruption/rejoin boundaries, using post-rejoin ticks first and, when necessary, the immediately pre-partition ticks. In 49.6\% of seeds in each case, the Adaptive high-rate count exceeds the number that can be placed after rejoin alone, so a pre-partition high-rate window is required for exact count matching. Mean assigned budgets are 6,810.5 pair opportunities in Case A and 6,813.6 in Case B, with exact equality among the three timing policies within every case and seed.

\subsection{CRN validation and recovery definition}
The validation layer records hashes for the proposal trace, broadcast-channel trace, and gossip-opportunity trace. Across all 2,000 case--seed bundles (1,000 seeds in each of two cases), the three-policy structure, exact budget equality, proposal-trace equality, broadcast-trace equality, and gossip-opportunity-trace equality all passed without failure. This establishes that the timing comparison changes when matched synchronization opportunities are allocated rather than silently changing the proposal or channel realization through policy-dependent random-number consumption.

Recovery in the CRN timing analysis is defined independently of simulator event density. A one-second monitor tick evaluates all-head agreement. ``First sustained recovery'' is the first event for which all heads remain equal for 30 consecutive one-second observations. The analysis additionally records final agreement at the horizon, the fraction of post-rejoin monitor ticks in all-head agreement, whether agreement is lost after first sustained recovery, the number of agreement-loss episodes, and the first time from which agreement persists until the horizon. Factorial and robustness screening analyses elsewhere in this Supplement use the event-based detector and are not numerically mixed with the CRN recovery-time estimates below. Table~\ref{tab:crn-equal-budget-cells} reports the cell-level outcomes for the three equal-budget timing policies.

\begin{table}[H]
\centering
\caption{Cell statistics for the CRN equal-budget timing experiment (1,000 matched seeds per case; Full head rule). Assigned budget is the pre-drop policy counter and is exactly equal across the three policies within every seed.}
\label{tab:crn-equal-budget-cells}
\scriptsize
\resizebox{\textwidth}{!}{%
\begin{tabular}{llrrrr}
\toprule
Case & Timing & Final agreement & Recovery mean (s) & Post-rejoin agreement fraction & Mean assigned budget\\
\midrule
A 50/50 & Adaptive & .832 & 90.52 & .8278 & 6810.5\\
& DisruptionWindowMatched & .809 & 88.18 & .8292 & 6810.5\\
& FixedMatched & .794 & 104.94 & .7832 & 6810.5\\
\midrule
B 80/20 & Adaptive & .834 & 84.28 & .8332 & 6813.6\\
& DisruptionWindowMatched & .816 & 83.11 & .8339 & 6813.6\\
& FixedMatched & .787 & 96.39 & .7889 & 6813.6\\
\bottomrule
\end{tabular}%
}
\end{table}

The corresponding paired policy differences and exact McNemar tests are reported in Table~\ref{tab:crn-equal-budget-effects}.

\begin{table}[H]
\centering
\caption{Paired CRN timing effects. Final-agreement intervals are 95\% paired-bootstrap intervals in percentage points, recovery intervals are paired-bootstrap intervals in seconds, and $p$ values are exact McNemar tests on discordant final-agreement pairs.}
\label{tab:crn-equal-budget-effects}
\scriptsize
\resizebox{\textwidth}{!}{%
\begin{tabular}{llrrr}
\toprule
Case & Comparison (first minus second) & Final agreement difference (pp) & Recovery difference (s) & McNemar $p$\\
\midrule
A & Adaptive $-$ FixedMatched & +3.8 [+2.5,+5.1] & -14.43 [-17.24,-11.75] & $1.62\times10^{-9}$\\
A & Adaptive $-$ DisruptionWindowMatched & +2.3 [+1.4,+3.3] & +2.33 [+1.27,+3.52] & $1.55\times10^{-6}$\\
A & DisruptionWindowMatched $-$ FixedMatched & +1.5 [+0.1,+2.9] & -16.76 [-19.56,-14.02] & .0489\\
\midrule
B & Adaptive $-$ FixedMatched & +4.7 [+3.4,+6.1] & -12.11 [-14.92,-9.38] & $1.18\times10^{-12}$\\
B & Adaptive $-$ DisruptionWindowMatched & +1.8 [+0.9,+2.8] & +1.16 [+0.48,+1.92] & $1.21\times10^{-4}$\\
B & DisruptionWindowMatched $-$ FixedMatched & +2.9 [+1.6,+4.3] & -13.28 [-16.05,-10.62] & $5.70\times10^{-5}$\\
\bottomrule
\end{tabular}%
}
\end{table}

The CRN results preserve the main equal-quantity timing observation but narrow its interpretation. Adaptive remains clearly better than approximately uniform FixedMatched timing: final agreement is higher by 3.8 and 4.7 percentage points, mean first sustained recovery is 14.4 and 12.1 s faster, and post-rejoin agreement fraction is higher by about 0.044 in Cases A and B. However, DisruptionWindowMatched explains most of the recovery-time effect. It recovers 2.3 and 1.2 s faster than Adaptive and has essentially the same post-rejoin agreement fraction, while Adaptive retains a smaller final-agreement advantage of 2.3 and 1.8 points. The supported conclusion is therefore that disruption-focused temporal concentration is substantially better than approximately uniform allocation at the same assigned quantity; the oracle-derived contextual trigger adds a modest end-of-horizon agreement benefit in this simulator but does not provide a general recovery-speed advantage over the deterministic disruption-window control.

\subsection{Durability diagnostics}
All 6,000 CRN runs reach the first sustained-agreement event defined by 30 consecutive one-second observations, but every run subsequently loses all-head agreement at least once while proposals continue. Mean agreement-loss episodes range from 30.9 to 32.8 across the six case-policy cells. Final stable agreement is available only in runs that finish in agreement, with availability equal to the final-agreement rates (.787--.834); among those runs its mean start time is about 1,169 s after rejoin, close to the simulation horizon. These diagnostics show why first sustained recovery is an event-time measure rather than permanent convergence. Final agreement and post-rejoin agreement fraction are therefore reported as complementary durability endpoints.

\section{End-to-end low-budget no-context reference}
\begin{figure}[H]
\centering
\includegraphics[width=.82\linewidth]{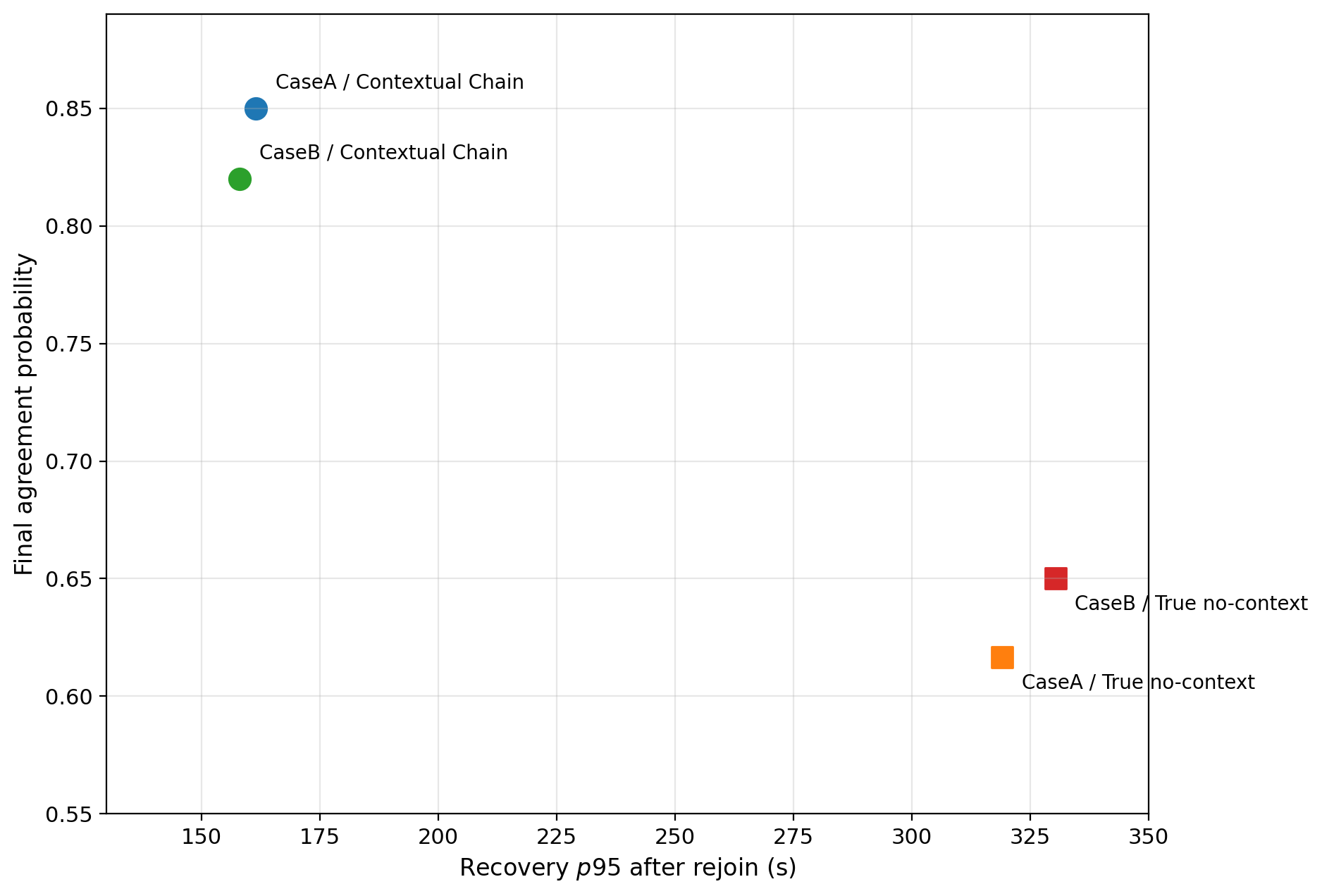}
\caption{End-to-end low-budget no-context reference. The comparison is confounded with synchronization resources and is therefore not used for causal attribution.}
\label{fig:true-no-context}
\end{figure}

A separate end-to-end ``true no-context fixed-budget'' configuration removes epoch-labelled ordering, branch-score guidance, quarantine, adaptive control, and the higher synchronization budget. Because this changes several factors simultaneously, the comparison cannot isolate a contextual effect and is used only as an end-to-end low-budget reference. Figure~\ref{fig:true-no-context} shows this confounded end-to-end comparison.

In that separate experiment, Case A reported final agreement 0.850 with recovery $p95=161.50$ s for the coupled configuration and 0.617 with $p95=319.05$ s for the low-budget no-context reference. Case B reported 0.820 / 158.05 s and 0.650 / 330.30 s, respectively. These numbers should not be compared directly with the 1,000-seed main table because the experiment used a separate seed set and configuration wrapper.

\section{Partition-ratio robustness}
The same qualitative provisioning pattern appears for a 90/10 split in the synthetic noisy setting. Final agreement was 0.614 for \texttt{NoQ}, 0.579 for \texttt{Q\_only}, 0.839 for \texttt{Gossip\_only}, and 0.843 for \texttt{Both}. Corresponding recovery $p95$ values were 336.10, 348.05, 146.00, and 151.00 s. This supports only the qualitative statement that the low- versus higher-synchronization separation is not specific to 50/50 or 80/20.

\section{Longer horizon}
\begin{figure}[H]
\centering
\includegraphics[width=.88\linewidth]{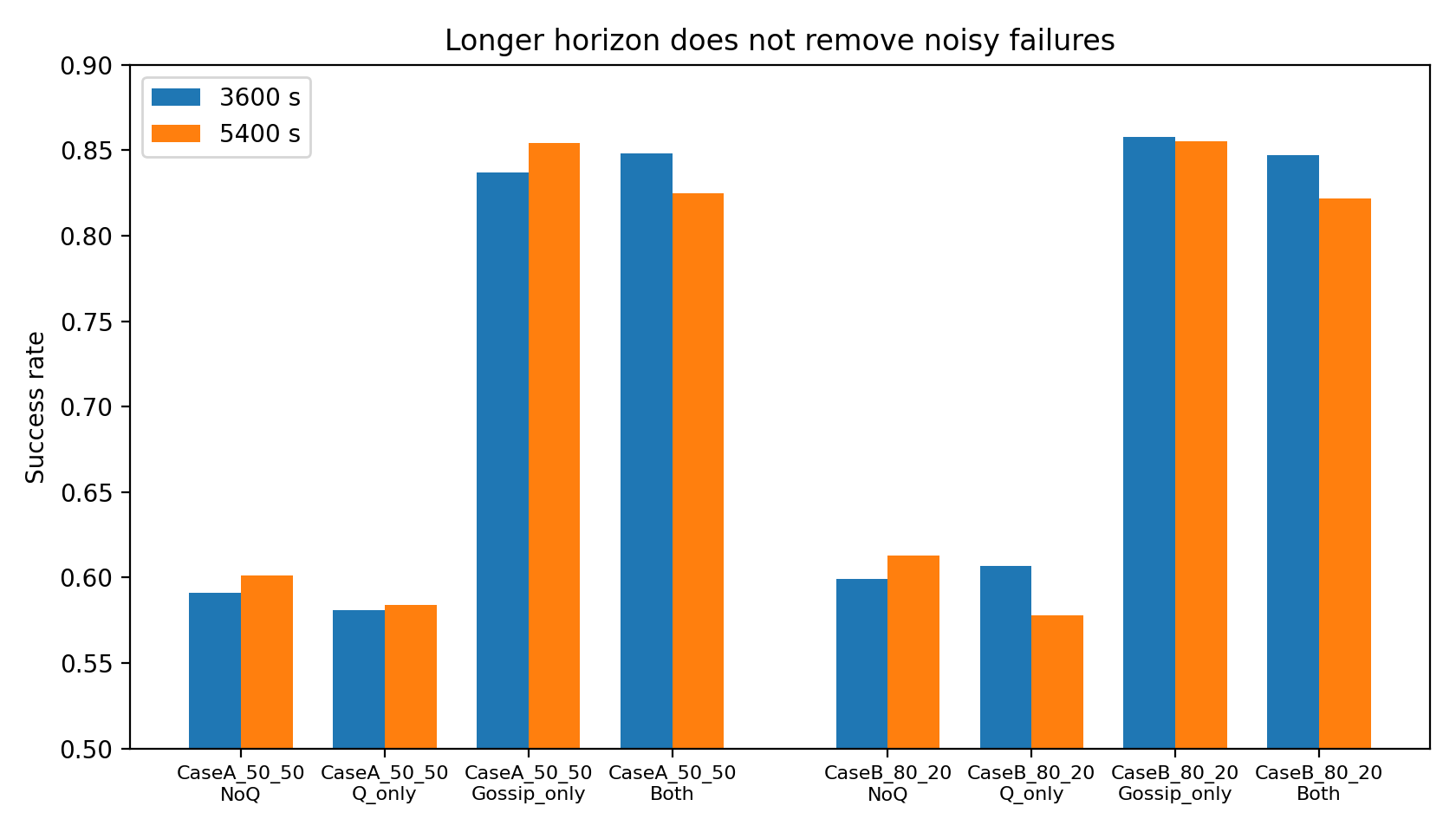}
\caption{Longer-horizon comparison. The figure provides a robustness check and does not establish eventual convergence.}
\label{fig:longer-horizon}
\end{figure}

Extending the total simulation horizon from 3,600 s to 5,400 s does not remove the low-budget gap, as shown in Figure~\ref{fig:longer-horizon}.

For Case A synthetic-noisy, \texttt{NoQ} changes from 0.591 to 0.601 final agreement; \texttt{Gossip\_only} is 0.854 and \texttt{Both} 0.825 at the longer horizon. For Case B, \texttt{NoQ} changes from 0.599 to 0.613, \texttt{Gossip\_only} is 0.855, and \texttt{Both} 0.822. Thus merely waiting longer does not make the one-pair regime resemble the higher-synchronization regime in these experiments.

\section{Parameter sensitivity}
\begin{figure}[H]
\centering
\begin{subfigure}{.48\linewidth}\includegraphics[width=\linewidth]{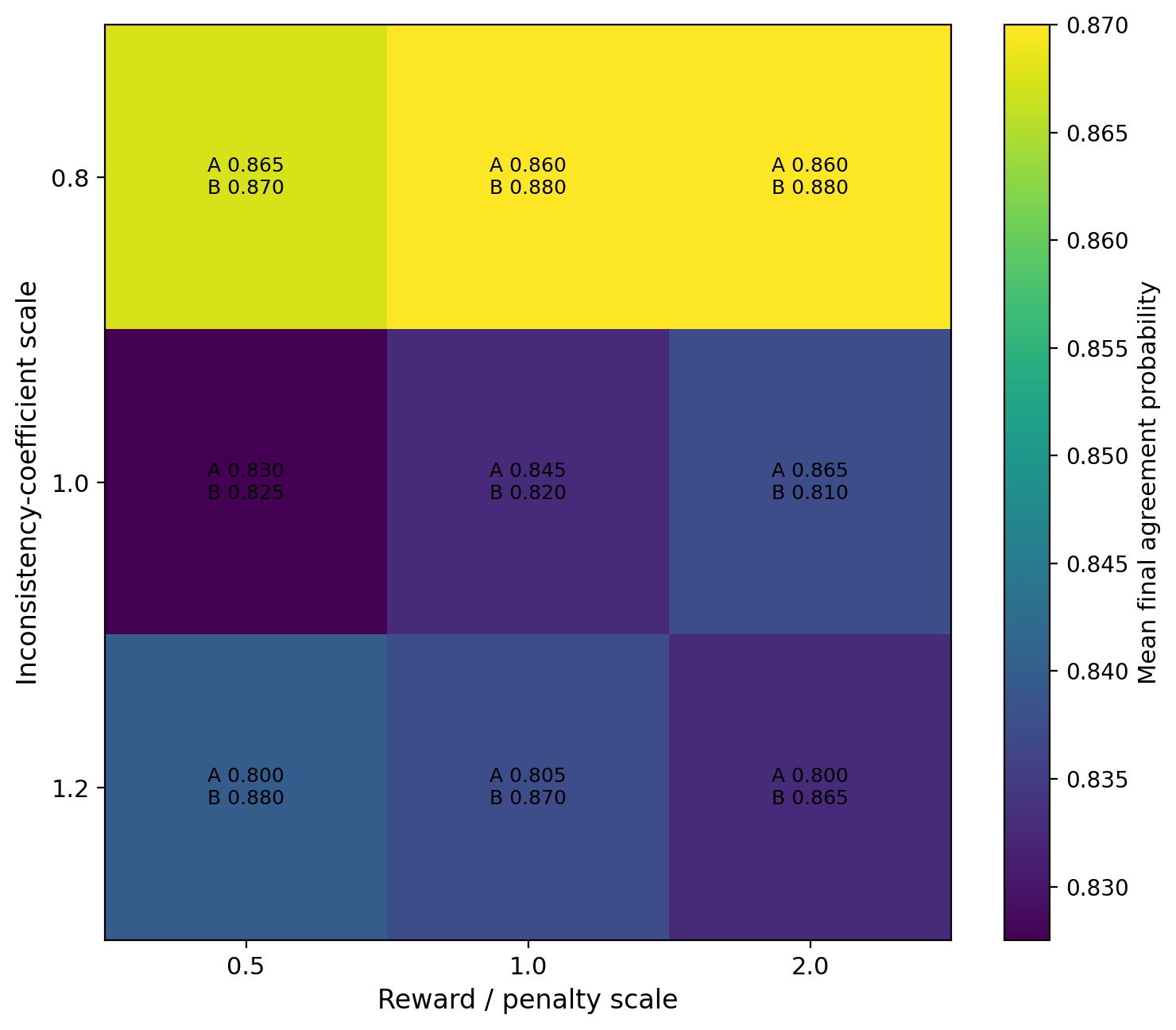}\caption{Final agreement.}\end{subfigure}\hfill
\begin{subfigure}{.48\linewidth}\includegraphics[width=\linewidth]{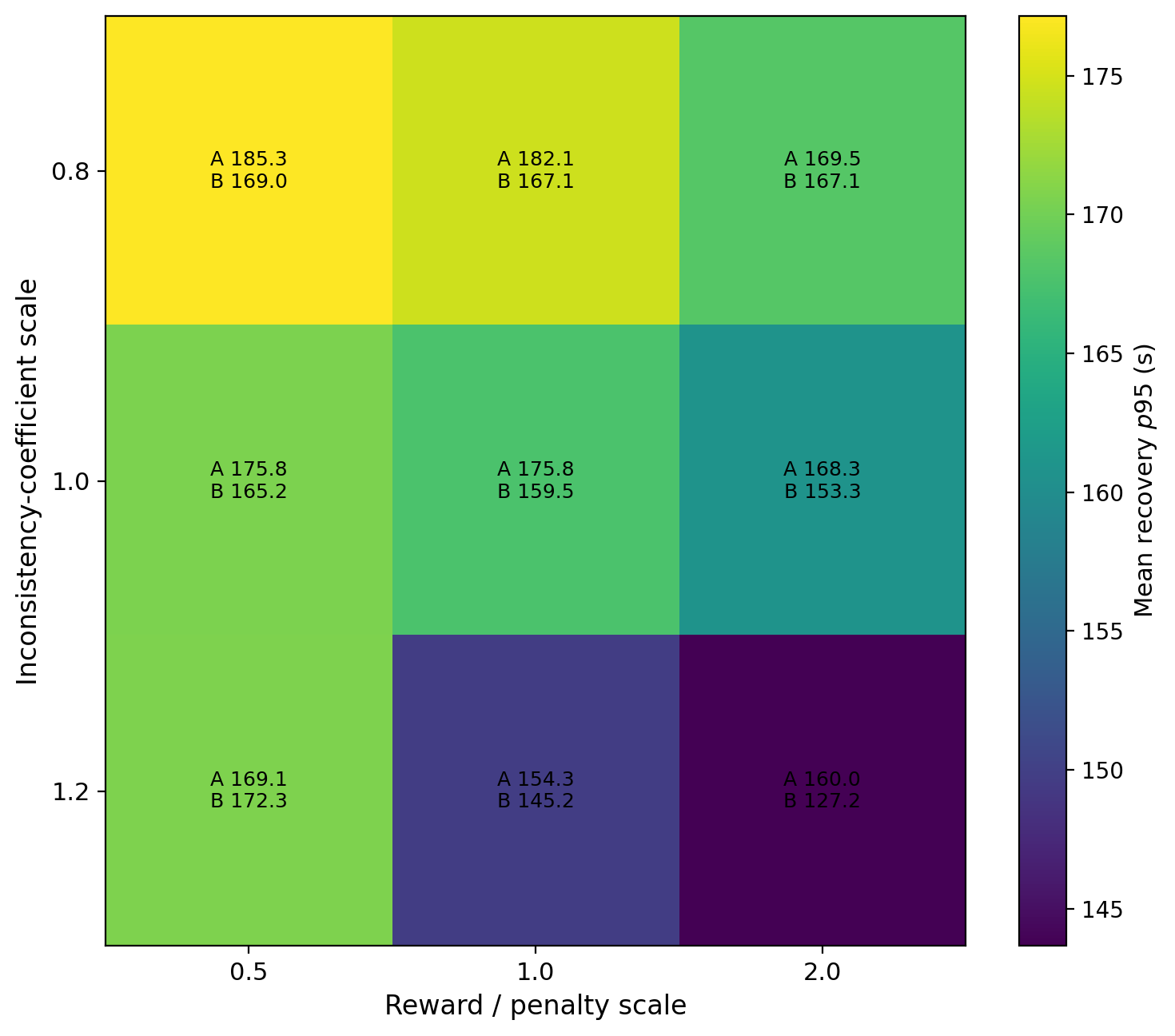}\caption{Recovery $p95$.}\end{subfigure}
\caption{Joint-parameter sensitivity study. The analysis does not establish theoretical optimality of the default constants.}
\label{fig:parameter-sensitivity}
\end{figure}

The sensitivity grid jointly scales the coefficients in the inconsistency formula by $\{0.8,1.0,1.2\}$ and the reward/penalty magnitudes by $\{0.5,1.0,2.0\}$. It provides a local robustness check rather than a full ablation of $\alpha$, $T_{on}$, $T_{off}$, $S$, and all reputation constants. Figure~\ref{fig:parameter-sensitivity} summarizes the resulting final-agreement and recovery-$p95$ patterns.

The explored grid did not overturn the principal low- versus higher-synchronization pattern. It is not, however, a complete one-factor-at-a-time or global sensitivity study. In particular, the EMA factor, hysteresis thresholds, off-streak length, checkpoint/epoch length, and reputation update constants require broader analysis before robustness can be generalized.

\section{Checkpoint-mode diagnostic}
The simulator supports height-based and time-based epoch labels. In the main height-based sticky-free mode, the epoch level is a deterministic monotone function of height and therefore does not add an independent fork-choice signal. The time-based diagnostic changes that relationship and serves as a robustness check; Table~\ref{tab:checkpoint-mode} reports the corresponding outcomes.

\begin{table}[H]
\centering
\caption{Sticky-free height-based versus time-based epoch mode in the main noisy suite.}
\label{tab:checkpoint-mode}
\scriptsize
\begin{tabular}{llcccc}
\toprule
Case & Variant & Success (height) & Success (time) & $p95$ height & $p95$ time\\
\midrule
A & \texttt{NoQ} & .591 & .575 & 344.05 & 352.15\\
A & \texttt{Q\_only} & .581 & .606 & 365.10 & 366.05\\
A & \texttt{Gossip\_only} & .837 & .860 & 152.05 & 153.10\\
A & \texttt{Both} & .848 & .838 & 172.00 & 169.05\\
B & \texttt{NoQ} & .599 & .611 & 329.10 & 334.35\\
B & \texttt{Q\_only} & .607 & .621 & 338.05 & 330.05\\
B & \texttt{Gossip\_only} & .858 & .867 & 154.05 & 148.05\\
B & \texttt{Both} & .847 & .838 & 166.00 & 159.00\\
\bottomrule
\end{tabular}
\end{table}

The main conclusion does not depend on treating height-based epoch level as a substantive checkpoint signal. The main Article therefore describes the rule as ``epoch-labelled height'' and does not assign an independent checkpoint role to that label.

\section{Scaling diagnostics}
\begin{figure}[H]
\centering
\begin{subfigure}{.48\linewidth}\includegraphics[width=\linewidth]{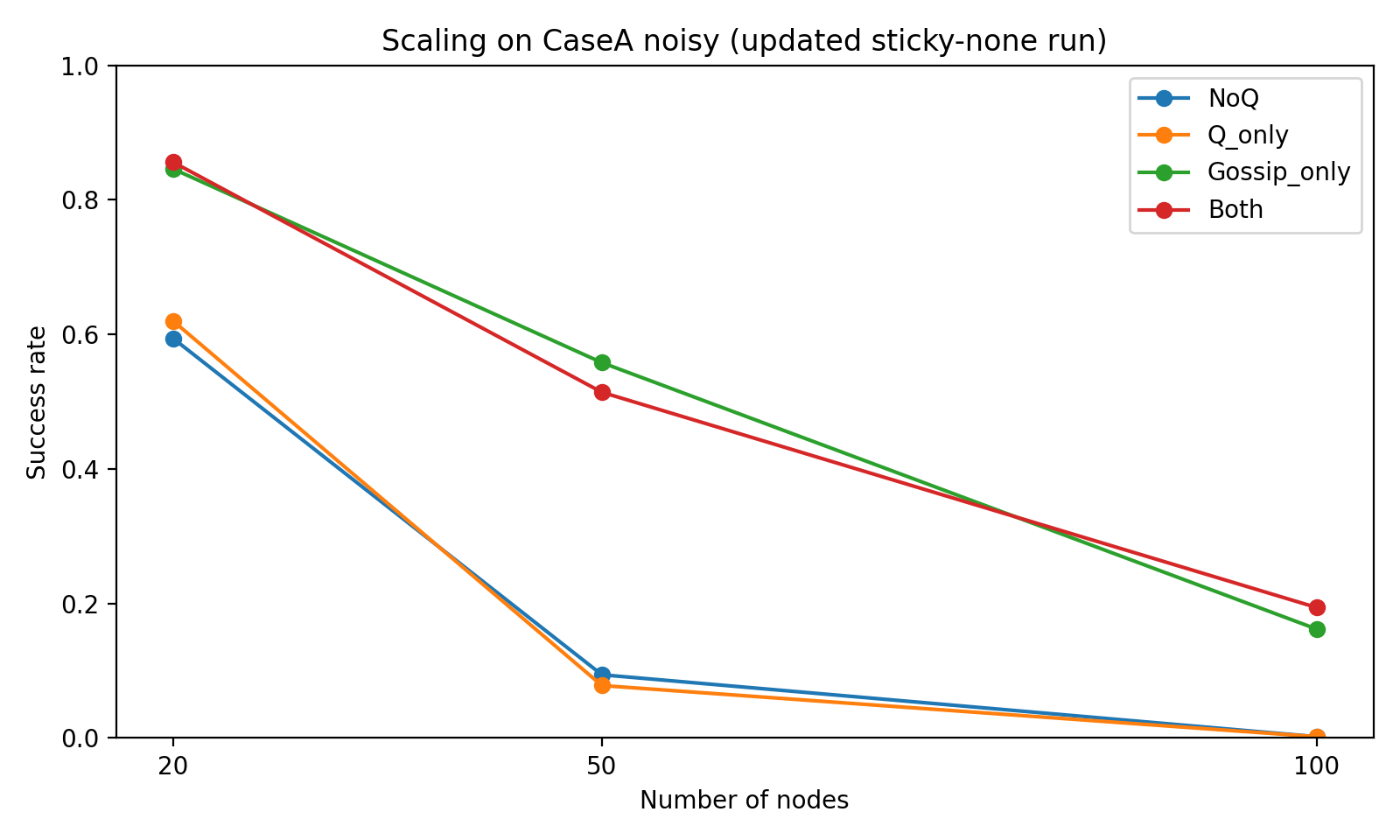}\caption{Final agreement.}\end{subfigure}\hfill
\begin{subfigure}{.48\linewidth}\includegraphics[width=\linewidth]{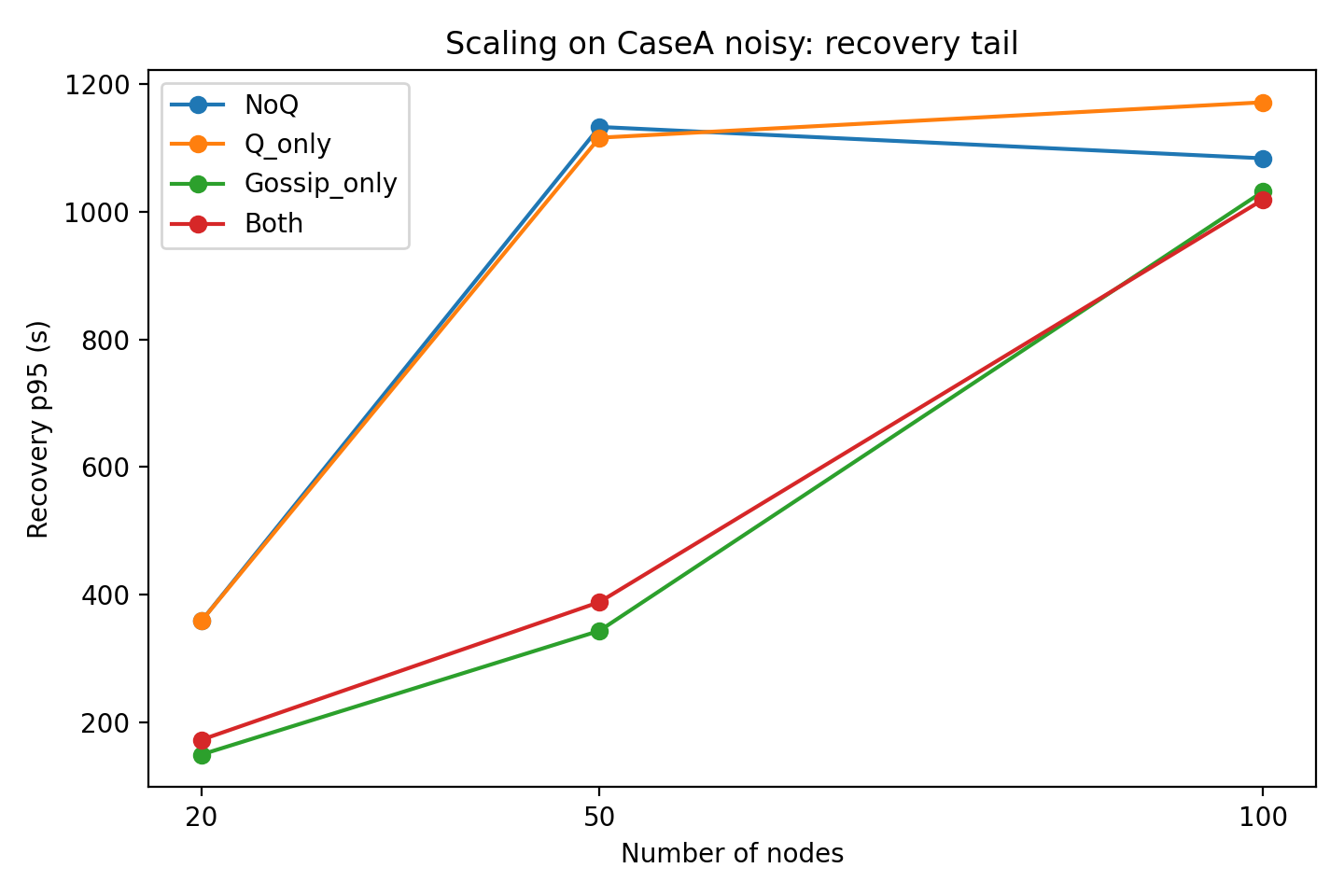}\caption{Conditional recovery $p95$.}\end{subfigure}
\caption{Case A scaling diagnostics. The main Article reports recovery-by-horizon and RMST to account for unrecovered runs.}
\label{fig:scaling-diagnostics}
\end{figure}

Figure~\ref{fig:scaling-diagnostics} provides the scaling diagnostics. Because total gossip pair budgets are not normalized by $N$, these plots are interpreted as fixed-budget provisioning stress tests rather than intrinsic scaling laws.

A future scaling study should compare constant per-node, per-edge, or per-block synchronization budgets; extend the budget sweep to $N=100$; model finite-capacity links; and evaluate topology-aware peer selection, clustered synchronization, or relay/super-node structures. The present paper makes no claim that increasing an all-to-all gossip budget indefinitely is sustainable.

\section{Context-coverage resource model}
\label{sec:coverage}
\subsection{Purpose}
This section presents a \emph{budget-sensitive context-coverage model} that separates the longer-term cryptographic motivation from the systems evidence. It asks a resource-accounting question only: if a recovery trace contains several simultaneously plausible contexts, how much contextual state would an observer need to retain to answer challenges that may refer to those contexts?

The observer is intentionally strong with respect to communication: it sees the active-context set produced by the underlying recovery trace and is constrained only by a context-storage budget. It is not subjected to an additional noisy link, bandwidth cap, or hidden packet stream. This isolates storage coverage and should not be interpreted as a realistic network attacker model.

\subsection{Exact challenge model}
Let $\mathcal C_t$ be the set of $m_t$ active contexts at time $t$. Suppose an observer can retain at most $b$ contexts and a challenge samples $k$ distinct active contexts uniformly without replacement. If the observer has selected $r=\min(b,m_t)$ contexts, the direct context-coverage probability is
\[
P_t(\mathrm{cover}\mid m_t,b,k)=
\begin{cases}
0,& r<k,\\[4pt]
\dfrac{\binom{r}{k}}{\binom{m_t}{k}},& r\ge k.
\end{cases}
\]
Thus the sharp budget effect in the context-coverage experiment is a direct consequence of combinatorial coverage once $m_t$ grows. The formula follows directly from the coverage model; the simulation supplies the time-varying $m_t$ generated by a partition/rejoin trace.

This clarification is important. The experiment does not demonstrate a cryptographic property, does not model an optimal adversary, and does not prove that an attacker must literally keep all contexts. It only measures the contextual multiplicity that a later formal construction would have to exploit.

\subsection{Reported exploratory results}
The Case A resource experiment uses budgets $b\in\{1,2,4,8,16\}$. Table~\ref{tab:coverage} reports the resulting context-coverage quantities using resource-accounting terminology.

\begin{table}[H]
\centering
\caption{Exploratory context-coverage results for Case A. Memory values are MiB in the simulator's context accounting. ``Required peak'' and context-count columns describe the underlying recovery trace and therefore do not depend on the observer's storage budget.}
\label{tab:coverage}
\scriptsize
\resizebox{\textwidth}{!}{%
\begin{tabular}{rrrrrrrrr}
\toprule
Budget & Challenge coverage & Rejoin coverage & End coverage & Stored peak & Required peak & Peak ratio & Peak contexts & Rejoin contexts\\
\midrule
1  & .568 & .000 & .840 & .67 & 5.28 & 7.89 & 9.29 & 8.57\\
2  & .717 & .036 & 1.000 & 1.33 & 5.28 & 7.89 & 9.29 & 8.57\\
4  & .835 & .219 & 1.000 & 2.31 & 5.28 & 7.89 & 9.29 & 8.57\\
8  & .984 & .818 & 1.000 & 4.42 & 5.28 & 7.89 & 9.29 & 8.57\\
16 & 1.000 & 1.000 & 1.000 & 5.28 & 5.28 & 7.89 & 9.29 & 8.57\\
\bottomrule
\end{tabular}%
}
\end{table}

\begin{figure}[H]
\centering
\begin{subfigure}{.48\linewidth}\includegraphics[width=\linewidth]{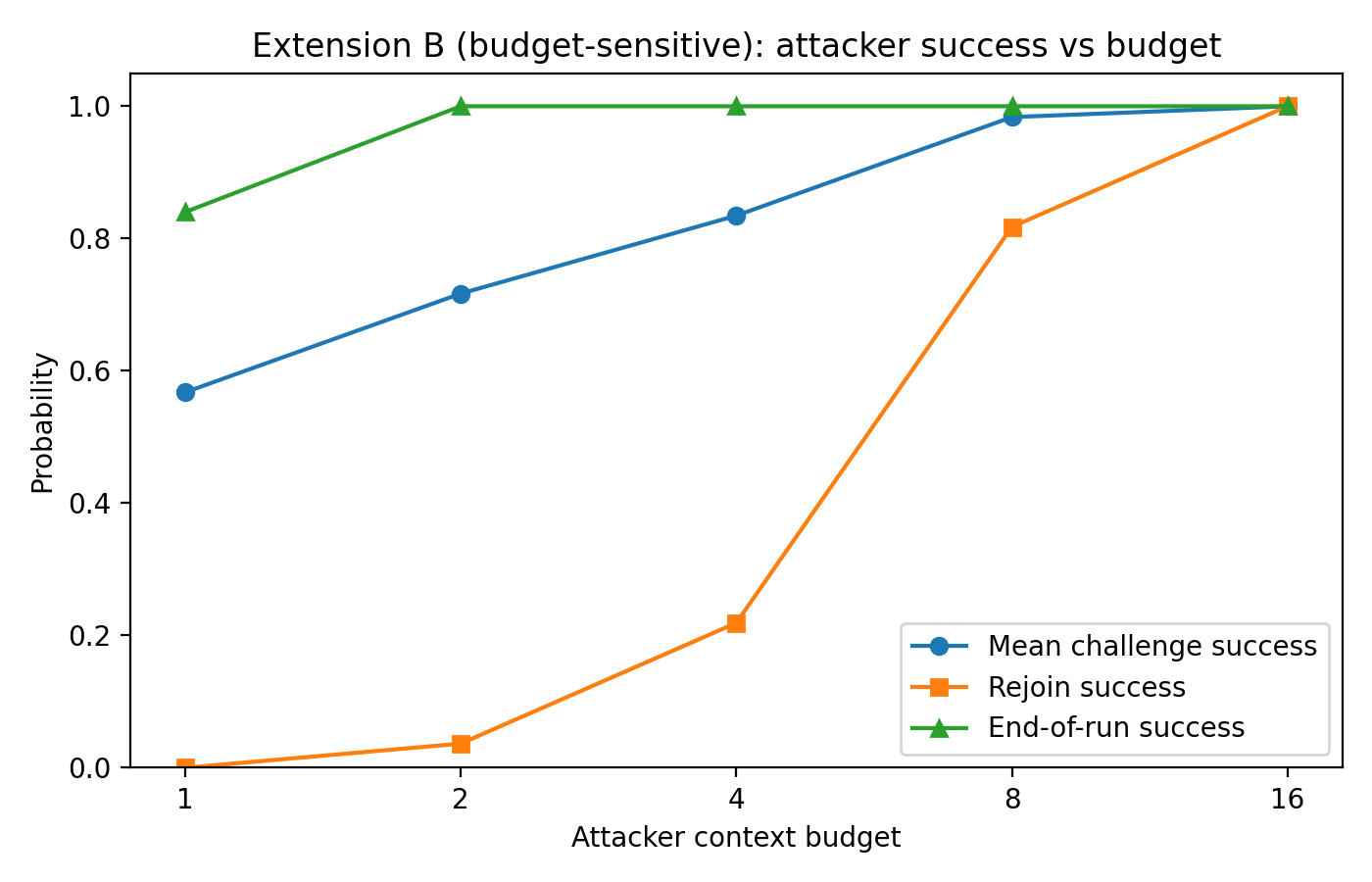}\caption{Coverage success versus budget.}\end{subfigure}\hfill
\begin{subfigure}{.48\linewidth}\includegraphics[width=\linewidth]{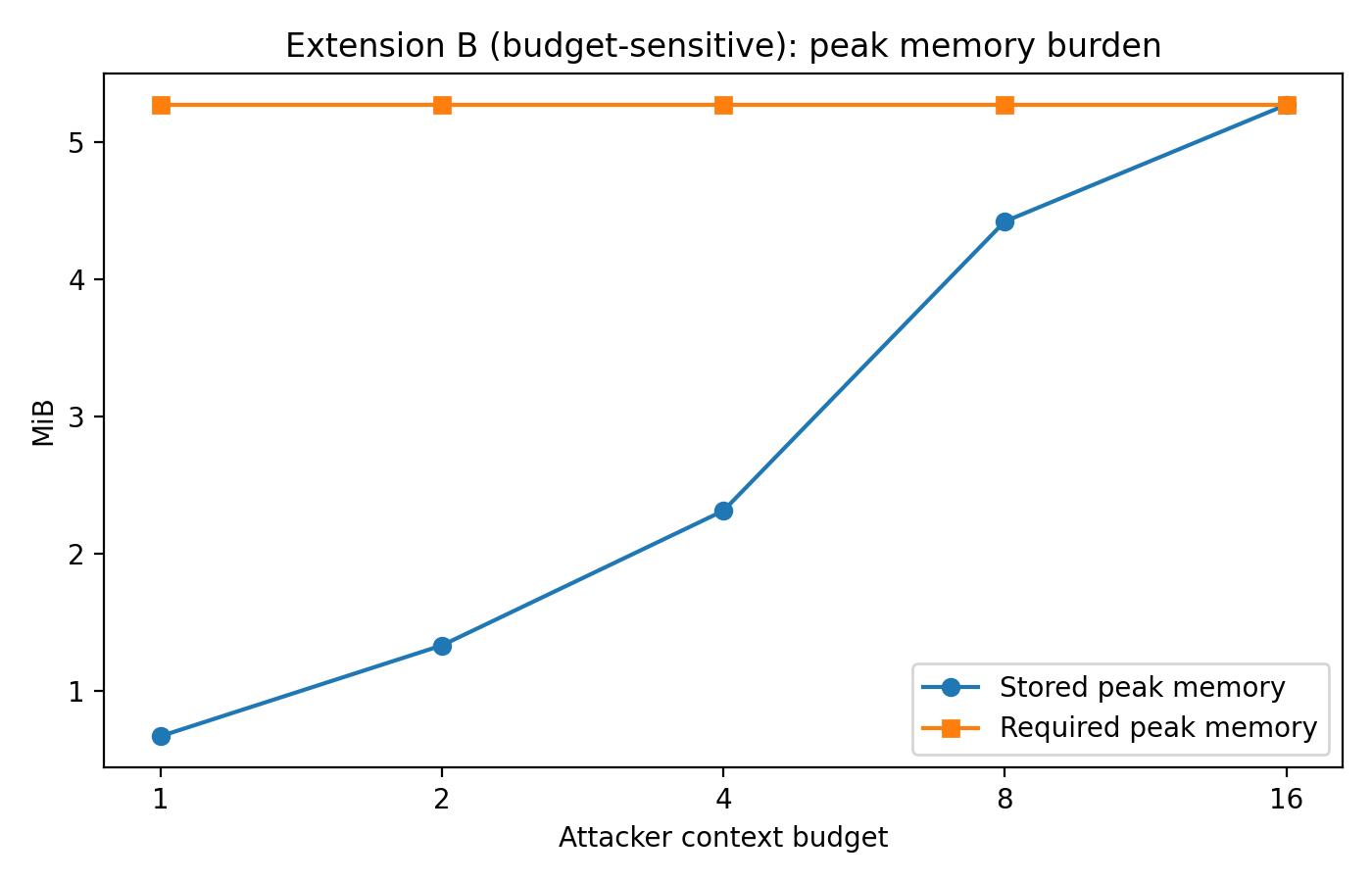}\caption{Stored versus required context memory.}\end{subfigure}
\caption{Exploratory context-coverage outputs. These are resource-accounting results, not cryptographic security evidence.}
\label{fig:coverage-outputs}
\end{figure}

The budget-independent quantities ``Required peak'', ``peak contexts'', and ``rejoin contexts'' are properties of the same underlying active-context trace. Changing $b$ changes how many of those contexts the observer stores, not how many the trace generates. This is why those columns are constant across budget rows. Figure~\ref{fig:coverage-outputs} visualizes the corresponding coverage-success and stored-versus-required-memory patterns.

The mean active-context multiplicity peaks around rejoin (9.29 at peak and 8.57 at rejoin in this Case A run family), while the settled end state is close to one context. The corresponding full-coverage peak was 5.28 MiB versus 0.67 MiB for the single-context honest-state proxy in this accounting, a ratio of 7.89. In a separate Case B check, reported peak and rejoin context counts were 9.59 and 8.57, with required peak 5.74 MiB versus 0.71 MiB (8.13$\times$). These values motivate further study of context-induced state asymmetry but do not establish that a real adversary has no better strategy.

\subsection{What a cryptographic follow-up would require}
A cryptographic paper would need to define honest and adversarial interfaces; specify how contexts are committed; define challenge distributions and freshness; model observation, erasure, and preprocessing; state the adversary's computational and memory resources; and prove completeness and soundness. DRG/Merkle constructions, proof-of-space techniques, bounded-storage authentication, or other concrete primitives could be considered, but none is implemented or claimed here. Suffix-targeted and checkpoint-attested challenges are therefore listed as future alternatives rather than simulated evidence.

\section{Radio/transport deployment matrix}
The simulator is deliberately above the radio layer. Table~\ref{tab:radio} explains how a later validation should map common IoT technologies rather than treating them as interchangeable.

\begin{table}[H]
\centering
\caption{Qualitative deployment mapping. These are not simulator-calibrated claims.}
\label{tab:radio}
\small
\begin{tabular}{p{2.1cm}p{3cm}p{6.4cm}}
\toprule
Technology family & Typical systems role & Parameters that must replace the current synthetic model\\
\midrule
Zigbee / IEEE 802.15.4 & Short-range low-power mesh & Measured topology, hop-dependent loss/delay, MAC retries, frame limits, duty cycle, per-hop energy\\
Bluetooth Low Energy & Short-range low-power links / star or mesh profiles & Connection intervals, MTU, retransmission behavior, topology/profile, energy per exchange\\
LoRaWAN & Long-range LPWAN & Data rate, MTU, duty-cycle limits, time-on-air, acknowledgement mode, fragmentation/SCHC, gateway/satellite contact windows\\
Sigfox-class LPWAN & Long-range small-payload LPWAN & Message/payload constraints, duty-cycle/access rules, asymmetric downlink availability, measured loss/contact schedule\\
\bottomrule
\end{tabular}
\end{table}

The main design implication is that the adaptive repair budget must be expressed in technology-specific cost units (airtime, packets, energy, or contact opportunity), not simply in abstract gossip pairs.

\section{Reproducibility notes}
The reproducibility package is self-contained for the controlled analyses. It includes the simulator used for the screening analyses, the separated-stream CRN simulator, trace-validation script, three-policy equal-budget runner, all 6,000 per-run outputs, matched-schedule diagnostics, paired statistical tables, provenance, SHA-256 checksums, and reproduction instructions. The tagged software release (\texttt{v2.1-minor-revision-2026-09-06}) is archived on Zenodo at DOI 10.5281/zenodo.22523507. The archived release contains the CRN materials used for the timing claims and permits regeneration of the reported statistics from the per-seed outputs.

\end{document}